\documentstyle[12pt,aasms4,epsfig]{article}

\begin{document}

\title{ Ultraviolet Spectrophotometry of Variable Early-Type Be and B stars
Derived from High-Resolution IUE Data}

\author{Myron A. Smith$^1$ }
{\footnotesize $^1$STScI/CSC, Space Telescope Science Institute, 3700 
San Martin Dr. Baltimore, MD 21218 and Catholic University of America,
Washington, D.C.;  ~~Email:~ msmith@stsci.edu } 

\begin{abstract}

   High-dispersion {\it IUE} data encode significant information about 
aggregate line absorptions that cannot be conveniently extracted from 
individual stellar spectra. Herein we apply a new technique 
in which fluxes from each echelle order of a short-wavelength {\it IUE} 
spectrum are binned together to construct low-resolution spectra 
of a rapidly varying B or Be star. The division of binned spectra obtained 
during a ``bright-star" phase by spectra from a ``faint-star" phase 
leads to a ratioed spectrum which contains information about the mechanism
responsible for a star's variability. The most likely candidate mechanisms 
are either the periodic or episodic occultations of the star by ejected 
matter or a change in photospheric structure, e.g. from pulsation.
We model the variations caused by these mechanism by means of model 
atmosphere and absorbing-slab codes. 
Line absorptions strength changes are rather sensitive to physical conditions 
in circumstellar shells and ``clouds" at temperatures of 8,000--13,000~K, 
which is the regime expected for circumstellar structures of early B stars.

   To demonstrate proofs of concept, we construct spectral 
ratios for circumstellar structures associated with flux
variability in various Be stars: 
(1) Vela\,X-1 has a bow-shock wind trailing its neutron star companion; 
at successive phases and hence in different sectors, the wind exhibits 
spectrophotometric signatures of a 13,000\,K or 26,000\,K medium, 
(2) 88\,Her undergoes episodic ``outbursts" during which its UV flux fades, 
followed a year later by a dimming at visible wavelengths as well; the 
ratioed spectrum indicates the ``phase lag" is a result of a nearly gray 
opacity that dominates all wavelengths 
as the shell expands from the star and cools, 
permitting the absorptions in the visible to ``catch up" to those in the UV,
and (3) $\zeta$\,Tau and 60\,Cyg exhibit periodic spectrum and flux changes, 
which match model absorptions for occulting clouds but are actually most
easily seen from selective variations of various resonance lines. In addition, 
ratioed UV spectra of radial and large-amplitude nonradial pulsating stars 
show unique spectrophotometric signatures which can be simulated with model 
atmospheres. An analysis of ratioed spectra obtained for a representative 
sample of 18 classical Be stars known to have rapid periodic flux variations 
indicates that 13 of them have ratioed spectra which are relatively featureless
or have signatures of pulsation. Ratioed spectra of three others in the 
sample exhibit signatures that are consistent with the presence of 
co-rotating clouds.

\end{abstract}
\keywords{stars: Be -- stars: circumstellar matter -- stars: pulsation }

\section{Introduction}
\label{intro} 

  The cause or causes of the rapid light and spectrum variability of the 
``classical Be stars"\footnote{Classical Be stars as defined herein are 
post-zero age main sequence, non-interacting-binary B stars for which the 
Balmer lines have been observed in emission. 
This emission is generally as evidence of sporadic 
ejections of mass which form a flattened circumstellar disk.}
is a longstanding problem with no convincing universal solution at 
hand (see Baade \& Balona 1994, Balona 1995,
Smith 1999, Balona 2000, Baade, 2000).
Both radial and nonradial pulsations 
(hereafter, ``NRP") are endemic to this region of the H-R Diagram, but the 
physical attributes and manifestations of short- and long-period modes can 
be quite different. In many respects the latter modes are not well understood. 
Likewise, generally indirect evidence has been advanced that magnetic 
dissipation processes are somehow responsible for this rapid activity, even 
though magnetic fields have not been directly detected in nearly all Be stars 
examined. An indicator of magnetically-triggered expulsions of matter leading
to Be ``outbursts" could be the presence of magnetospheric structures 
similar to the co-rotating tori over He-rich Bp stars 
(Shore \& Brown 1990). These tori (herein ``clouds") absorb stellar flux as 
they occult the stellar disk and produce variability in broad-band UV light 
curves. If the structures are optically thick they can also produce variations 
in spectral line profiles of a rotating star that are similar in appearance
to those produced by NRP. A broad consensus has formed (e.g., Smith 2000b, 
Baade 2000) that either or both mechanisms can play a role in ejecting mass 
from the surfaces of these stars, but even if both of them are operative 
their relative roles are 
unclear. To establish these roles requires that the mechanism for the 
variability be correctly assigned. Except for a relative handful of 
multiple-periodic Be stars, determining the variability mechanism is 
not always straightforward. Statistical studies (e.g., Balona 1990) suggest
that the dominant periods of monoperiodic Be stars in visible-band light 
are consistent with the stars' rotational periods, although the latter 
quantities are generally not well known. On the other hand, the
variability periods are also in agreement with the range expected
for long-period NRPs arising from g-modes.
  
  In this paper we pose the question of whether candidate physical causes
of the variability, most likely magnetospheric clouds or pulsations, can be 
determined in a Be star which is variable on a rotational timescale (see
also Smith 2000a). In phrasing this question, we make the 
assumption that circumstellar clouds should be {\em cooler} than the line 
forming region of the nearby photosphere and thus should cause preferential
absorptions among low-excitation metallic lines as they pass between the star 
and the observer. We will address this issue by introducing a 
spectrophotometric technique tailored to high-resolution, far-ultraviolet
echellograms of the {\it International Ultraviolet Explorer} ({\it IUE}) 
satellite processed by {\it NEWSIPS} software.\footnote{NEW Spectral 
Imaging Processing System (see Garhart et al. 1997).}
This technique proceeds by binning all fluxes within each of 60 echelle orders 
to form an equivalent low-resolution spectrum. We then ratio this spectrum with 
another spectrum observed at a time when the flux of the star is different, 
that is, by dividing spectra 
from a ``bright-star" phase(s) with those from a ``faint-star" phase(s). 
The ratioed spectra so formed contain residual signatures
of the variability mechanism because of the increased absorptions of an
intervening medium or, in the case of pulsation because the temperature 
distribution or velocities in the photosphere are expected to be modified
during a pulsation cycle.
The resulting ratioed spectra are simulated by dividing a synthesized 
photospheric spectrum into a spectrum computed with absorptions from
a slab of material in the line of sight.

  Recently, Smith \& Groote (2001) applied this spectrophotometric 
technique to 
evaluate physical conditions (temperature, turbulence, area, and column density)
in magnetic torus-shaped clouds forced into corotation over two magnetic B 
stars, $\sigma$\,Ori~E and $\beta$\,Cep. They found that the
weak metallic lines which are the principal absorbers in the far-UV
are formed in column densities of about 10$^{23}$ particles cm$^{-2}$
and temperatures of 11,500--13,000~K. Because absorptions of metallic
lines and continuous absorptions of hydrogen and H$^{-}$ generally 
increase with decreasing temperature, these authors pointed out that 
spectrophotometric detections technique are most sensitive to low cloud 
temperatures (8,000--10,000~K). However, there are other ways to infer 
the existence of a warmer material in an intervening medium. 
Smith \& Groote found that the principal UV 
resonance lines of spectra of these stars are very strong, suggesting 
that the clouds must also have a warm component. 
As previous investigators found before them, Smith \& Groote
reported that the UV profiles from the clouds associated with He-strong
Bp stars maintain their shapes and strengths at a given phase over at
least several years. In this paper, we ask whether putative clouds 
associated with Be stars could have these same general properties.

\section{Reduction and Analysis of Archival IUE Data }
\label{dataan}

\subsection{Data Reduction}

   The {\it IUE} echellogram data for this study were the high-dispersion 
spectral files from the Short Wavelength Prime (SWP) camera. The echellograms 
were downloaded from the MAST\footnote{Multi-Mission Archive at Space Telescope 
Science Institute, in contract to NASA.} data archive. 
In most cases, and wherever possible, large-aperture observations were used. 
The data selected for comparison were obtained generally observations made 
within a day of one another, typically in an intensive monitoring program. 
These data selections essentially avoid errors due to long-term 
degradations in the camera as well as changes in the reference point 
in the aperture used to recenter the stellar image. 
One exception to our selection practice is for a proof-of-concept 
demonstration which we make below for 88\,Her, the variations of which 
occur usually over a timescale of years. In separate work we have found 
that {\it NEWSIPS} calibrations are imperfect. For example, incomplete 
corrections in the {\it NEWSIPS} calibrations result in a residual systematic 
flux error of -1\% yr$^{-1}$ at most {\it SWP}-camera wavelengths;
this correction is somewhat larger for wavelengths below $\lambda$1250 
(Smith 1999a). These dependences can cause slowly varying 
time-dependent errors in the derived background fluxes of high-dispersion 
spectra, with resulting errors in the zero-point of the flux scale which 
affect net spectra at short wavelengths. Fortunately, these effects are
generally unimportant over short timescales covered in intensive monitoring 
campaigns such as those we utilized principally in this study. In addition,
ratios of spectral fluxes tend to be insensitive to such errors. Despite 
these grounds for optimism, an important part of our data conditioning
procedure was the monitoring of the depths of highly saturated 
resonance lines such as the C\,IV and Si\,IV doublets, Lyman$~\alpha$, 
and Si\,III at $\lambda$1206 and $\lambda$1298. Because these lines are 
already completely saturated in the cores, the core depths cannot vary 
for astrophysical reasons, such as changes in atmospheric temperature 
or the added absorptions from circumstellar material.
For spectra taken close together in time, the core depths of these lines 
remained constant to within $\pm{2}$\% of the continuum level for nearly all 
the exposures in any one group. Thus, we adopted this value as the uncertainty 
of the background levels in the program spectra. For wavelengths above 
$\lambda$1500 these errors drop to about half this value. 

    Our numerical analysis of our data proceeded by means of a computer
program which first loops through all echelle spectra for a given star and 
tabulates the core depths of saturated lines mentioned above. Attributing 
fluctuations of these depths to variations in 
the background level, we interpolated the implied zero-point offsets 
for orders between the resonance lines and extended them to these factors 
for short and long wavelength echelle orders. 
Our code then computes corrections to the 
net fluxes\footnote{Herein we treat exclusively net fluxes rather than 
``absolute calibrated" fluxes in order to avoid unnecessary 
spectrum-to-spectrum errors in the ripple-correction process. The net
fluxes of an echelle order provide a natural wavelength bin unit because
they rise from low values at one end of the order and drop to low values
again at the other end.} 
by applying the (nearly unitary) ratios of the average core depths for 
all spectra to the core depths of each individual spectrum.
The code next sums all valid fluxes in each order (that is, it omits 
pixels with bad data quality flags) and creates a low-resolution spectrum
from them. This process was repeated for each spectrum in a time series, 
producing a stacked array of low-resolution, high signal-to-noise spectral 
ratios. The effective resolution of the binned spectrum is given by 
the echelle's free spectral range, which is typically about 20 Angstroms.
   
  The final phase of the data manipulation consists of computing a 
pseudo-measure of the ultraviolet continuum (hereafter, ``UVC"), which is 
the average of the summed fluxes of four orders extending over the range 
$\lambda\lambda$1800--1905. We then relied upon a variability timescale, 
which is a period determined from ground-based or UV studies and is usually 
obvious in our {\it IUE}-derived flux curves as well. High-resolution
spectra obtained at times when the star was UVC-bright were binned, 
co-added, and divided by the sum of spectra when the UVC flux was faint. 
This manipulation permits the computation of the ratio:

\begin{equation}  
F_{obs}(\lambda) = 1 - f_{min}(\lambda)/f_{max}(\lambda)~.
\end{equation}    

\noindent We will use the term ``ratioed spectrum" as short-hand
in this paper to refer to this equation. 
Depending on the astrophysical mechanism causing the variability, 
this spectrophotometric ratio can be related either to a theoretical ratio 
of two photospheric spectra or to one minus the occulted spectrum to the 
unocculted photospheric spectrum.

  Errors in the ratioed spectra depend on several factors. One of the most
important of these is the loss of sensitivity at short wavelengths. 
Both Smith, Robinson, \& Hatzes (1998) and Smith \& Groote (2001) determined 
r.m.s. errors of $\pm{0.8}$\% at long wavelengths which increased to 
$\pm{1.5}$\% at short wavelengths using binned flux-ratioing of {\it IUE} data. 
We have examined the r.m.s. fluctuations of a few ostensibly nonvariable 
stars and found values averaging $\pm{1.1}$\% for wavelengths above
$\lambda$1300. These estimates are based on both internal (point-to-point)
variations and comparisons of ratioed spectra from data obtained from
different groups of observations. For this study we have relied on two
new approaches to address systematic errors, which we describe as follows.

  In the first approach, we selected a pair of echellograms for the B3Ve 
star $\alpha$\,Eri. These observations (SWP~55873 and SWP~55889) were 
obtained 0.7 days apart, which is about half the principal
1.26 day variability cycle discovered by Balona,
Engelbrecht, \& Marang (1987; see also Leister et al. 2000). 
We constructed a ratioed spectrum both
from {\it NEWSIPS}- and {\it IUESIPS}-processed data. The background flux
solutions derived by these processing systems are altogether different. 
Although the {\it NEWSIPS}-derived values are usually more accurate (but 
typically slightly low), they were derived by sampling non-local as well as
local fluxes on the echellogram. In contrast, {\it IUESIPS} backgrounds 
were determined only from local fluxes and tend to
be high. Figure~\ref{msfig1} shows spectral ratios from both extractions.
We take their differences to be a generous estimate of errors resulting
from flux extractions. 
Note that both curves start near unity at long wavelengths and increase
almost monotonically toward short wavelengths. A ``spike" at Lyman~$\alpha$ 
in the ratioed spectrum resulting from the fluxes of both constituent 
spectra being low and hence producing an uncertain ratio. 
At long wavelengths the mean ratios of the 
{\it IUESIPS} and {\it NEWSIPS} spectra are nearly identical, and the 
difference between them is typically $\pm{1.5}$\%. At shorter wavelengths 
these errors increase until at $\lambda$1160 
the curves become meaningless. The important result here is that the 
curves are mutually consistent, suggesting that the low-resolution 
spectral ratios are insensitive to the method of flux extraction. 

  To evaluate the effects of background flux errors, we repeated our
manipulation of the same pair of {\it NEWSIPS} spectra, but this
time we lowered the background by an assumed error which is linear in 
decreasing wavelength and normalized it to -2\% of full continuum at
1250 Angstroms. The dashed curve in Fig.~\ref{msfig1} exhibits the
result of dividing this modified spectrum by the original one -- that is,
the result of the perturbed background flux levels. This curve
again increases toward shorter wavelength and also causes a faint
spike at Lyman~$\alpha$. 
In all, although the amplitude of these features is small, the
spectrum-to-spectrum deviations in background level can potentially 
could produce second-order errors at short wavelengths.

   As a second approach to evaluating systematics in ratioed spectra, we
constructed an artificial ``knife-edge test" by comparing pairs of spectra 
of a nonvariable star. In one case one spectrum was obtained with the star 
within the aperture, and in the other case it had begun to drift to 
the aperture edge. 
We selected two pairs of small aperture spectra for this purpose
for the stars 16~Lac and $\eta$~Ori (SWP~05858 \& 05860
and SWP~22163 \& 22158). These observations were obtained during times 
when the stars appeared otherwise constant in brightness. 
Inverting the ratios to represent an artificial brightening, we found that 
the ratioed spectra exhibited slopes of -20\% and -2\% with respect to 
their mean flux ratios minus one. Thus, this is a nearly flat but
slightly red response.
Both ratioed spectra showed r.m.s. fluctuations of about 1\%, which is 
consistent with the r.m.s. values quoted above.

\subsection{Computation of Simulated Spectra and Spectral Absorptions} 

  To model the observed spectral ratios, we used a suite of codes written 
by I. Hubeny and collaborators to compute synthesized ratios of stellar 
spectra (in the case of pulsation) or absorptions of the stellar spectrum 
(in the case of an occulting medium). To mimic our actions on the observed
data, we first divided the spectra one by the other and then binned them 
to the effective free spectral range of the {\it SWP} camera. 
For the case of pulsation, in which one photospheric 
spectrum is divided by another, the comparison to the observations is 
conceptually straightforward because computed and observed ratios are nearly 
unity. For the case of absorbing clouds the computed absorptions are 
normalized to the stellar continuum. Then, a weak absorption spectrum ratio 
will be some fraction between zero and one.

   The first of the Hubeny codes, {\it SYNSPEC}, is a spectral line 
synthesis program developed for input non-LTE and/or line-blanketed model 
atmospheres (Hubeny, Lanz, \& Jeffery 1994). We used standard 
LTE model atmospheres by Kurucz (1993) as input to {\it SYNSPEC}. 
{\it SYNSPEC} is embedded within an IDL wrapper (Hubeny 1996) to facilitate 
the calculation of spectra for a variety of atmospheric parameters.
For our program we ran models with log~g = 4, a microturbulence of 
5 km~s$^{-1}$, and no rotational broadening.
Another available option in {\it SYNSPEC} is to modify the run of  
atmospheric temperature, T($\rho$x). We used this option to mimic variations 
in the temperature gradient which occur during a radial pulsation.

To simulate the effects of a cloud on the composite spectrum,
we used the Hubeny {\it CIRCUS} program (Hubeny 1996, Hubeny \& Heap 1996). 
This code was written to compute line absorptions and/or emissions 
of a gas cloud situated either in front or off the limb of a reference star.
These absorptions comprise the so-called ``iron curtain" in the ultraviolet.
{\it CIRCUS} requires the user to specify physical cloud parameters such as
temperature, density, geometry, composition, microturbulence, column depth,
and areal coverage factor. The computations are performed at the same
wavelengths as for the synthesized photospheric spectrum. {\it CIRCUS} 
can accommodate clouds with as many as three separate sets of conditions,
sizes, and positions. However, in this work we considered mainly homogeneous
clouds. In its solution of the radiative transfer equation {\it CIRCUS} 
computes line emission and absorptions separately. These options can be
treated as ``switches" and permits the investigator to evaluate the two 
effects separately. 
The cloud temperatures we consider are generally less than 20,000~K.
For this temperature range reemission in both the resonance transitions
and weak iron-curtain lines is negligible.

  {\it CIRCUS} proceeds by consulting a Kurucz (1990) line library 
of atomic absorption parameters and computing an opacity spectrum for 
a user-specified temperature, composition (solar), and electron density. 
The spectra were computed at a 
spacing of 0.01 Angstroms, thereby resolving most UV lines in the Kurucz
line library. The optical depth in each line is determined from the 
computed opacity spectrum and the input column density. 
The surface of the star is divided into a grid and the local intensity
spectrum is evaluated at each grid point, thereby taking into account 
effects of foreshortening and limb darkening. 

  The density and geometrical coverage factors of the assumed cloud are 
additional necessary parameters in our analysis. For our initial models we
started with a trial volume density of 10$^{12}$ cm$^{-3}$. 
This estimate is based on the 
probable cloud densities of 10$^{11}$--10$^{12}$ cm$^{-3}$ that appear to 
prevail in the rotating magnetic B stars (e.g. Smith \& Groote 2001). 
For simplicity we ran models with a full (100\%) areal coverage factor. 
The actual coverage factor for any given case may be obtained by scaling the
full-coverage models to the observed level of the ratioed spectra (minus one).
The positions and amplitudes of individual absorption peaks in the observed 
spectral ratio can then be used to determine the cloud temperature and 
column density, respectively.

 {\it CIRCUS} includes a provision for various doppler effects, 
including stellar rotation. The program computes the net doppler 
velocity between each projected element of the cloud and the background 
star along the observer's line of sight. For a co-rotating cloud the
velocities along the line of sight are the same, so the net shift is zero. 
Another relevant factor in determining cloud absorptions is the 
microturbulence. 
However, Smith \& Groote (2001) found that velocities less than 20 km~s$^{-1}$ 
produce similar cloud spectra, so we used this value as a default.

\section{Results for General Models}

\subsection{Homogeneous Clouds}
\label{homog}

  Over a broad range of possible cloud temperatures, the density of
metallic metallic lines increases with decreasing temperature in nearly 
all ultraviolet and visible band regions. 
For a reference cloud model of T$_{cloud}$ $\approx$ 13,000~K and a 
particle density N$_{H}$ = 3$\times$10$^{-12}$~cm$^{-3}$, the sum of the 
hydrogen bound-free absorption and electron scattering coefficients is 
about 8$\times$10$^{-25}$ cm$^{2}$. 
For this model we tabulated the number of local maxima in computed 
absorption spectrum and found a total of some 26,000 unblended lines. 
The median strength of these lines is only 65\% as strong as the continuous 
opacity, so the majority of the lines becomes optically thick 
at almost the same column density as the underlying opacity continuum. 
In the optically thin limit the wavelength binning of the absorption 
spectrum to $\sim$20\AA~ generates amorphous ``features" due to local line 
aggregates which can arise above the
continuous absorption by a factor of up to 2--3. A detailed inspection of 
the line strength distribution in the $\lambda\lambda$1150--2000 wavelength 
region shows that the median strength is typically higher than the 
lower-envelope of the absorptions by only tens of percent. 
Thus, except in the case of strong lines (Lyman~$\alpha$, resonance lines
of Si\,III, Si\,IV, and C\,IV), the binned spectrophotometric features are 
nearly as optically thin as the absorption windows between them. This fact 
explains why the calculated spectra are insensitive to the microturbulence 
velocity parameter $\xi_{t}$ over a range of values. 
For intermediate column densities the optically thin features diminish
in amplitude and gradually lose their identities.
As one increases the column still further, the absorptions finally 
become optically thick at all wavelengths. For cloud temperatures of 8,000~K
and 13,000~K, this happens for column densities of about 1.5$\times$10$^{23}$ 
and 3$\times$10$^{24}$ cm$^{-2}$, respectively. At this point the fraction of 
absorbed flux increases to about 80\% to 90\%, respectively, and
the spectrum shape becomes almost flat. 

  Although the morphology of the absorption spectrum depends on the 
contrast with the underlying photospheric spectra, our {\it CIRCUS} models 
suggest that the detailed features are well preserved over a typical 
range in effective temperatures of $\pm{3000}$~K for T$_{eff}$ $>$
13,000~K.
Figure~\ref{msfig2} exhibits absorption spectra computed for a variety of
indicated cloud temperatures, a column density of 1$\times$10$^{23}$ cm$^{-2}$,
N$_{H}$ = 3$\times$10$^{11}$ cm$^{-3}$, $\xi_{t}$ = 20 km~s$^{-1}$, and 
coverage factor of 1.  
Because the postulated clouds co-rotate with the star, the results are the 
same for nonrotating and rotating star inputs. We have also run models 
of rapidly rotating stars and with intervening clouds having no doppler
component along the line of sight, i.e., the case of quasi-static {\em shells.}
The primary departures from the co-rotating cloud models are that resonance
lines exhibit slightly added absorptions because their saturations are 
lifted by the velocity difference along the line of sight.  

  Changes in cloud temperature greatly modify the shape of the 
ratioed spectrum, particularly for values less than 11,000~K.
This is a combined result of the increased
hydrogenic absorption and a shift from third to second-ionization stages of
chiefly the Fe-group elements. Consider especially how cool temperatures
affect strong lines: the most prominent feature arises from the
resonance Si\,III line at 1206~\AA~ and Lyman$~\alpha$, which dominate two 
adjacent wavelength bins. Other sharp features in the binned ratioed spectrum
are caused by resonance lines of Si\,IV ($\approx$$\lambda$1400) and 
C\,IV ($\approx$$\lambda$1548-50), and from excited 
multiplets of C\,III ($\lambda$1176) and Si\,III ($\approx\lambda$1300).
All of these features are visible for temperatures $\ge$ 13,000~K. 

  Resonance lines are less visible for cool temperatures 
(T$_{cloud}$ $<$ 11,000~K). This is partly 
a consequence of the continuous opacity 
becoming stronger as hydrogen recombines. In addition, as these
lines disappear resonance lines of second-stage ions appear
at longer wavelengths, often beyond the {\it SWP} camera range. 
However, a few resonance lines of less
ionized species still play roles in the general rise of absorption for all 
cloud temperatures considered, particularly at wavelengths below $\lambda$1200. 
A good example is the confluence of Si\,II $\lambda\lambda$1190--1197 
lines which fall near a number of strong excited Fe\,III lines. 
In addition, at lower temperatures
the Si\,II lines saturate and form strong damping wings. 
These increased absorptions, along with the still-strong Si\,III 
$\lambda$1206 and Lyman~$\alpha$ lines form a strong composite absorption. 
When these features are binned with a group of nearby 
excited Si\,III and low-excitation Fe\,II lines, they produce
a general increase in absorption below Lyman~$\alpha$.

  At longer wavelengths several broad features form that provide
convenient diagnostics of temperatures in ratioed {\it SWP} spectra.
The most well known of these is the absorption centered 
at $\lambda$1900 (e.g., Bjorkman, Bjorkman, \& Wood 2000; hereafter ``BBW"). 
This broad feature
appears at a clouds temperatures between 10,000-17,000~K and attains 
its maximum strength at 15,000~K. The absorption is largely 
due to a variety of excited ($\chi$ = 8--10 eV) Fe\,III lines. 
For temperatures less than about 9,000~K the ionization of iron shifts from
Fe$^{2+}$ to Fe~$^{1+}$, causing an important Fe\,II absorption feature to 
appear at $\lambda\lambda$1600--1700. A secondary peak at $\lambda$1850 
is formed at cool temperatures largely from excited Fe\,II atoms. 
An additional weak peak centered near $\lambda$2050 is formed from an assorted
collection of excited second-ionization lines of iron-like atoms and Fe\,I 
resonance lines. Recently, Groote \& Smith (2001) were able to use
weak features near $\lambda$1650 and $\lambda$1900 in the binned spectral
ratios of the magnetic B stars $\sigma$~Ori~E and $\beta$~Cep to
determine temperatures in the outer regions of their co-rotating clouds.

\subsection{Nonradial and Radial Pulsations}
\label{puls}

 Nonradial pulsations on a star cause different regions to move both
horizontally and vertically, causing periodic modulations in atmospheric
temperature and pressure. An external observer viewing this activity 
will alternately view a dominance of first hot and then cool regions as the 
pattern moves across the projected stellar disk. By ratioing atmospheric 
fluxes from these regions, 
one obtains a spectrophotometric signature of the pulsations.
In Figure~\ref{msfig3}a we exhibit this behavior to first order by 
ratioing the synthesized spectra of static atmospheres having effective 
temperatures of 26,000~K and 22,000~K. This example clearly represents an
exaggeration of flux variations in most NRP stars, but it serves to emphasize 
the signatures of a simple atmospheric flux ratio. Even so, the only 
clear temperature signatures are weak flux-ratio elevations corresponding 
to the Lyman~$\alpha$ line and the Si\,III $\lambda$1300 multiplet. Unlike
the cloud absorption models, this ratioed spectrum displays no ``sharp"
feature at $\lambda\lambda$1100--1150. Small-amplitude fluctuations at long 
wavelengths are the result of an incomplete cancellation of a complex array 
of line strengths in the two constitutent spectra. Otherwise, this ratio 
may be roughly characterized as a monotonic function increasing with 
decreasing wavelength. Other authors (e.g., Peters 1991, Cugier, Daszynska, 
\& Polubek 1996, Peters 1998b) have exploited this dependence by fitting 
UV variations of NRP early-B stars to {\it IUE SWP} spectra.

  As the terms suggest, radial pulsations differ from nonradial pulsations
in that the locally imposed motions are vertical. 
As such, they are coherent over the entire stellar disk at any given phase. 
The local velocity amplitudes of the radial modes in a few $\beta$~Cephei 
stars can be very large, even exceeding the atmospheric sound velocity. 
Such motions typically modify the atmospheric structure by enhancing the 
temperature difference between the line and continuum formation regions. 
Occasionally, sharp absorption features appear at low negative velocities 
and then move redward through the line profile as a 
consequence of pulsational shock waves traveling through these atmospheres
(e.g., Campos \& Smith 1981, Smith 1983, Mathias, Gillet \& Crowe 1992). Phase 
delays are well known to occur for resonance lines formed in superficial 
atmospheric layers (e.g., Burger, de Jager, \& van den Oord 1982). We have 
pursued the Burger et al. finding by inspecting line fluxes of a number
of strong lines in the spectra of the large-amplitude pulsators BW\,Vul and
$\nu$\,Eri during their radial pulsation cycles. All the lines we investigated
showed a pronounced strengthening as the pulsation wave entered the line
forming region. Atmospheric simulations show that two ways of simulating 
these strengthenings is to lower the atmospheric boundary temperature 
(or in the more realistic formulation of non-LTE, one accounts for 
lowered excitation temperatures of lines due to the radiation boundary)
or to increase the velocity broadening (e.g., microturbulence) of the 
spectral lines. The temperature 
effect also can be inferred in the reddening of the colors of $\beta$\,Cep 
stars during the phase of maximum distension (relative to phase of compression).
Although dynamic model atmospheres do not exist for $\beta$\,Cep stars, 
we have utilized these facts to make a simplified approximation in creating 
a pair of modified model atmospheres by altering the temperature distribution 
T($\rho$\,x) for a T$_{eff}$ = 24,000~K, log~g = 4 Kurucz atmosphere to
mimic the effects of the passing wave in strengthening the ``temperature" 
gradient. The temperatures in the line formation regions ($\rho$\,x = 
10$^{-3}$--10$^{-2}$) of these models differ from the reference atmosphere 
by $\pm${400}~K, which is a typical temperature amplitude in the atmosphere
of a $\beta$~Cep star over the course of a pulsational cycle.
These distributions are displayed in Figure~\ref{msfig3}b. 
We also increased the atmospheric microturbulence from zero to 
15 km~s$^{-1}$ in the steep-gradient model in order to attempt to simulate 
the effects of turbulence following the passage of a shock wave (Stamford
\& Watson 1981). Synthesized spectra were then generated from these altered 
models and ratioed. The result is displayed in Fig.~\ref{msfig3}c. 
One can see from this that the results of perturbing the atmospheric 
distribution cause appreciable differences in the aggregate absorptions 
of spectral lines and therefore in the appearance of the two photospheric 
spectra. Thus, the ratioed spectra resulting from these simple descriptions 
of radial and nonradial pulsations (Fig.~\ref{msfig3}a) are quite different. 
Notice that flux contrasts at Lyman~$\alpha$ again produce a local spike
in the ratioed spectrum. However, a broad feature appearing
at $\lambda\lambda$1240--1375 has a more subtle explanation. 
The lines in this spectral region are comparatively sparse except for
several strong Si\,III and C\,II lines near $\lambda$1300.
The latter region is also nestled between the strong line Si\,III $\lambda$1206 
and Ly~$\alpha$ lines on one side and many stronger lines above $\lambda$1375. 
These lines respond sensitively to the steepening of the temperature gradient. 
Highly saturated lines in the spectrum, such as  $\lambda$1206, are sensitive 
to increases in microturbulence. The overall effect of this arrangement of 
strong lines is to create the illusion of a transparent window at 
$\lambda\lambda$1240--1380 punctuated by a sharp peak at $\lambda$1300. A
weak ``plateau" is often also present in the region between the Si\,IV and
C\,IV doublets, and this is due to a large assortment of Fe-group lines.
Two weak absorption features at $\lambda$1650 and $\lambda$2000 have similar 
explanations. 

  We emphasize that the particular atmospheric model described 
above in the radial pulsation simulation 
is simplistic and probably a nonunique. Furenlid et al. (1987) 
have suggested that line variations in BW\,Vul may be due in part to a
decrease in the atmospheric continuous opacity. We tested this possibility 
by comparing synthesized spectra obtained from 24,000~K models having 
log\,g = 4.5 and 3 with the same microturbulences as before.
We found that the ratioed spectra from these spectra 
show the same general spectrophotometric features as those 
shown in the Fig.~\ref{msfig3}c. Thus, we conclude that the
described spectral features are robust attributes which are produced by 
modifying the atmosphere in such a way as to strengthen the formation 
of most lines, as is indeed observed in {\it IUE} spectra.

\section{Analysis}   
\label{analy}

\subsection{Proofs of Concept: Intervening Slabs and Clouds}

  In this section we discuss in detail four variable B stars for which
independent evidence exists that circumstellar gas sometimes occults 
the star. All but the first of these examples are well-observed
classical Be stars which are
probably observed, respectively, through an expanding shell,
a density wake, or rapidly appearing and disappearing structure which could
be a magnetosphere.\footnote{For additional comparison to the Be stars, we 
also examined three interacting Be binaries for which the UV resonance lines
are known to show strong emission variations with phase (e.g., Peters \& 
Polidan 1984). Assuming several observations were present, these stars' 
spectra could not have be confused with those of Be stars or 
magnetic rotating B stars for several reasons. 
The continuuum fluxes show either negligible variations (e.g., CX\,Dra, 
HR\,2142) or variations so large that they must be caused by an eclipse of 
one stellar component (e.g., AU\,Mon). The resonance emission lines of these 
systems span a large range across the profile, including their blue wings, 
and are often morphologically distinct from one ion to another. A filling 
in of the C\,IV and Mg\,II profiles at some phases, which results from 
emission in a large volume of gas around the mass-receiving star, is not 
seen in spectra of the Be or magnetic B stars.} 
In each case, the ratioed spectrum displayed a pattern of spectrophotometric 
signatures which can be fit with {\it CIRCUS} models. These demonstrations 
show both the power and limitations to determinations of physical conditions 
of circumstellar matter from UV spectrophotometry.

\subsubsection{Vela\,X-1 ~=~ HD 77581 ~=~ 4U~0900-40}

   Vela\,X-1 is a high-mass X-ray binary composed of a B0.5~Ib primary and
a X-ray pulsar in a close, nearly-circular orbit. Because the components are
close ($a$ $\approx$ 1.6 R$_{*}$, P = 8.96 days) and the orbit is almost 
aligned to our line of sight 
($i$ $>$ 74$^{o}$; van Kerwijk et al. 1995), the neutron 
star transits the primary every orbital cycle.
The wind from the B star accretes onto a surface surrounding the pulsar,
permitting a detailed mapping of a complex wind environment. 
In the conventional picture (e.g., Kaper et al. 1993) 
wind particles emanating from the supergiant in this system 
form an accretion shock which trails the orbiting pulsar. 
This shock radiates at X-ray energies, resulting in the ionization
of an X-ray Stromgren sphere out to some distance. 
Because ultraviolet radiation is largely transparent in this region, the
accelerating force on the B-star wind particles nearly vanishes as they 
enter the Stromgren sphere, leaving them free to coast across the sphere 
at velocities they acquired before entering this region. 
As the particles leave the trailing edge of this sphere, they interact 
strongly with high velocity particles whose accelerations have not been 
interrupted. This interaction creates a second, spiral-shaped 
``photoionization wake" which runs from a point near the B-star's surface 
outward and behind the neutron star (Blondin et al. 1990). In these models 
the shocked region cools with distance from the Stromgren sphere, so 
material cools as one moves outward along the wake. 
The wake intersects the line of sight to the B star for perhaps a whole
cycle, so one expects to observe the cooled wake in UV resonance lines at 
phases much later than the transit of the pulsar (at $\phi$ = 0.5) and 
indeed just before the next transit. 
Kaper et al. (1993, 1994) have argued that Vela~X-1 is a case where these
complex interactions are readily observable. They point in
particular to line profile asymmetries in
a number of optical and ultraviolet resonance lines if one compares
them from phases just {\em after} (0.5 $<$ $\phi$ $\le$0.6) transit
compared to just {\em before} (0.4 $\le$$\phi$ $\le$0.5) this event.
These authors noticed that low-velocity features 
present at other orbital phases vanish at phases just after a transit and
attributed this to the absence of wind material
to an ionization change in the wind as it enters the photoionization wake.
By contrast, the accretion wake surrounding the pulsar is not likely to be
responsible for this variability because it is very small and also because 
it is too hot to absorb UV line flux.

  To investigate the prediction of Kaper et al., we have
formed ultraviolet quasi-continuum light curves from orders corresponding 
to $\lambda\lambda$1800--1905 from all 50 available {\it SWP} large-aperture 
spectra. We corrected the fluxes for the 1978 epochs by -7\% to bring their
values into agreement with late-epoch spectra; this correction is consistent
with the -1\%  yr$^{-1}$ noted above for degradation of long-term far-UV
{\it IUE} fluxes. The resulting light curve, 
displayed in Figure~\ref{msfig4}a, shows a slightly asymmetric $\approx$10\% 
dip at phases 0.46--0.70 centered at $\phi$ $\approx$ 0.52. 
Close inspection of the behavior of these absorptions reveals a different 
wavelength dependence just before and after light minimum. 
To make this point quantitatively, we have divided spectra during the
first (0.40 $<$ $\phi$ $<$0.46) and second (0.5 $<$ $\phi$ $\le$ 0.60)
halves of the pulsar transit by four spectra observed at 
phases prior to the pulsar 
transit ($\phi$ = 0.28--0.30). These spectra have the highest continuum 
fluxes (Fig.~\ref{msfig4}a) and are presumed to be least affected by the
effects of a photoionization wake. These ratios (minus one) are shown as 
Figure~\ref{msfig4}b and c. 
The ratioed spectrum in Fig.~\ref{msfig4}b exhibits long-wavelength 
absorptions, including a clear Fe\,III absorption bump at $\lambda$1900. 
As discussed in $\S$\ref{homog}, this feature
suggests a cloud temperature in the range 11,000--17000~K. For phases
just before transit, the ratioed spectrum in Fig.~\ref{msfig4}b can 
be reproduced quite well with a model having T$_{cloud}$ = 13,000~K, 
and a column density of 1$\times$ 10$^{23}$ cm$^{-2}$, all scaled to a 
coverage of $\approx$60\%. (For this 
model we used assumed parameters of T$_{eff}$ = 25,000~K, log\,g = 2 
for the computed underlying photospheric spectrum; see van Kerkwijk et al. 
1995). 
In contrast, the ratioed spectrum shown in Fig.~\ref{msfig4}c for 
post-transit phases is dominated by a broad hump at $\lambda$1550. 
This broad feature is not replicated by our moderate or 
cool-temperature models (cf. Fig.~\ref{msfig2}). However, models run  
with considerably higher temperatures show that the feature is due
to many aggregates of Fe\,III and IV lines. Our best-fitting match to the
ratioed spectrum was with a model with T$_{cloud}$ = 26,000~K. 
The absence of strong absorption features (other than from the C\,IV
doublet) at this temperature suggests a large column density, 
so we used a value of 3$\times$10$^{23}$ cm$^{-2}$. 
At this column density the absorption is large, and this must be
compensated to fit the observations by a relatively small coverage factor 
of 28\%. We should also point out that the sharp
dip in the observed spectral ratios at $\lambda$1400 (Si\,IV) 
is not necessarily a detriment to our shock-wake model. This feature is 
actually a {\em reflex} of warm wind material viewed at $\phi$ = 
0.3.\footnote{Judging from the P\,Cyg profiles of the Al\,III doublet at 
all phases,  
warm material is present in all lines of sight around the orbital cycle.} 
Overall, Fig.~\ref{msfig4}b is 
typical of the absorption spectrum one might expect from a moderately dense 
wind from an early-type B supergiant. In contrast, Fig.~\ref{msfig4}c 
suggests the presence of a local hot slab with a higher column density.
Finally, we remark that extracted He $\lambda$1640 line strengths from these
spectra (not shown) exhibit a clear increased absorption of this line lasting 
$\approx$0.35 cycles centered at $\phi$ = 0.5. This is the transit phase of 
the neutron star and indicates the presence of a hot source along this line
of sight. Putting this all together, these models argue for a wind 
the properties of which vary substantially with azimuthal position in the 
orbital plane. This is a good working description of the photoionization wake 
suggested by Kaper et al.

\subsubsection{88 Herculis = HD\,162732 }

 88\,Her is a late B-type classical B star (T$_{eff}$ = 15,000~K; Harmanec 
2000) which has undergone a series of long ``outbursts" over at least the 
last few decades. 
We define an outburst here by the appearances of Balmer and metallic 
``shell" lines accompanied by a decrease in continuum flux at UV and 
visible wavelengths (Harmanec et al. 1978, Doazan et al. 1982). 
It is generally believed that an outburst corresponds to the ejection of 
an initially opaque envelope which becomes optically thin as it slowly
expands over many stellar radii and cools. 
Be stars like 88\,Her typically decline in brightness at 
first and then gradually return to their pre-outburst levels. As the outburst 
cycle proceeds, the near- and far- ultraviolet spectra acquire low-excitation 
``shell" absorption lines of Fe\,III Fe\,II, Ni\,II, and similar ions
(Danezis  \& Theodossiou 1988, 1990). During 1976--1978
the UV brightness of 88\,Her faded by about one-third and by 1979 had
reattained its previous flux
(Barylak \& Doazan 1986). The visible-band flux faded as well, 
but its decline lagged by about by about a year behind the UVC curve.
In their compendium study of {\it UBV} light curves of Be stars, Pavlovski 
et al. (1997) showed that the visible flux of 88\,Her had nearly recovered 
to its former maximum by the early 1980's, only to fade once again in the 
mid-1980's. As with the previous cycle, the V-band fluxes lagged a year 
behind the U and B fluxes.

The archival high-dispersion {\it IUE} spectra cover the period 1981--92. 
The beginning of this interval coincided with a stabilization of the flux
from the previous outburst. Detailed examination of the {\it IUE} spectra 
shows that strong Fe\,III lines were present during this entire period, 
including the beginning of the outburst. Figure~\ref{msfig5}a exhibits the 
UVC light curve ($\lambda\lambda$1800--1905);
again, a +1\% yr$^{-1}$ correction has been applied to the fluxes.
The decreases in the flux show that another extinction episode occurring 
during the early 1980's. By 1987--8 the UV brightness had decreased by 40\%
and then brightened again during 1990--1992, virtually recovering to its 
initial maximum.

   We computed spectral ratios systematically as a function of epoch during
the mid- and late-1980's by using three spectra obtained near 1982.0 to 
represent the most unobscured state. 
The ratios depart appreciably from unity only after 1985 when the 
absorptions exceeded 10\%. In Figure~\ref{msfig5}b we display the ratioed
spectrum formed from spectra obtained in 1982.0 and 1986.0--.4. As one 
proceeds to the larger spectral ratios formed from 1987--8 data, 
corresponding to observations taken during the UVC minimum, the 
ratioed spectra become flatter and almost devoid of features. In principle, 
these could be easily confused with ratioed spectra of Be stars without
intervening matter (e.g., see Fig.\,9), except that they show a hint of
increased absorption near $\lambda$1650 and $\lambda$1900. More importantly,
they also show selective variations of resonance line strengths of Al\,III, 
Si\,III, and Fe\,III, a fact that demonstrates that cool material along the 
line of sight must be present at some times. A periodogram of the UVC light 
curve from one month to several years  
exhibits no peaks that might be present from gas streams between members 
of an interacting companion of long period. In particular, when folded over 
the 86.6-day orbital period suggested by Harmanec, Koubsky, \& Krpata (1974), 
the UVC fluxes show no correlation with phase. Thus, the UVC and 
line variations are not likely to be caused by binary interactions.
 
 Upon investigating constituent pairs of spectra in pre- and post-outburst 
groupings, we found that the various ratioed spectra we formed were
self-consistent and fit the ratioed spectrum with a {\it CIRCUS} model
having a low temperature and a fairly high column density. 
The best fits resulted with models with 
cool temperatures ($\approx$ 8,000~K) and high column 
densities. For these values, H$^{-}$ continuum 
opacity effectively dominates even the near-ultraviolet, so
the morphology of the absorptions is flatter than the lower column density
models depicted in Fig.\ref{msfig2}.
Figure~\ref{msfig5}b shows two fits to these for the necesssarily high 
column densities of 3$\times$10$^{23}$  and 6$\times$10$^{23}$ cm$^{-2}$. 
To obtain these fits we used coverage factors 
of 35\% and 28\% respectively. 
Since one might expect that a nearly optically thick cloud should be 
extensive, and thus should occult the star completely, we attempted to 
find models with 100\% coverage factors. However, we were unable 
to push coverage factors much higher (and the column densities lower) 
without producing substantially more prominent absorption features than 
are present in the data. Nonetheless, we believe that the fit to the 
observed features is quite good for wavelengths above $\sim$$\lambda$1250.

  Because the opacity increases with decreasing temperature,
the absorptions of hot gas embedded in a cool cloud can 
easily be concealed in the binned spectral ratios. 
Of course, this statement does not necessarily hold true 
for individual lines in a high-dispersion spectrum. 
In fact, an investigation of the Fe\,III line behavior also disclosed evidence
for the presence of warmer gas along the line of sight than the 8,000~K
component determined from the weak iron-curtain lines. This evidence
takes two forms. First, low-excitation lines of Fe\,III 
(and Al\,III) show occasional variations on a timescale of a few months. 
These variations do not correlate with the features in the UVC light curve in 
Fig.~\ref{msfig5}a. Second, the variations of these lines are largest among
moderate-excitation lines ($\approx$ 10\,eV). 
Fig.~\ref{msfig5}c shows a comparison of 
spectra in the $\lambda$1910 region obtained both in the UVC-minimum and 
recovery periods (i.e., at 1987.5 and 1990-2). Variations in 
the strengths of these lines are obvious. 
Absorptions for models with T$_{cloud}$ $\ge$ 13,000~K can easily produce 
changes in moderate-excitation lines, but low-excitation lines are too 
saturated already for the dense (or extensive) clouds to have much effect 
on them.

  According to the time histories of the Fe\,III line strengths, 
warm intervening gas was present present along the path to 
88\,Her during its entire outburst cycle during the 1980's. 
The amount of this gas varied at times over intervals
as short as a few months, perhaps as a result of fresh 
ejections of material. These additions cause the column length to vary 
by $\sim$3$\times$10$^{22}$ cm$^{-2}$ over intervals short 
compared to the total outburst cycle. From this evidence, rapid injections 
of warm gas seem to have expanded outward from the star 
and cooled during the
long outburst cycle. Such a picture leads naturally to an understanding
of the time lag of the visible flux after the UV minimum in 88\,Her during
a typical cycle:
initially the ejected plasma is warm and absorbs mainly in the far-UV. As 
the gas cools, the opacity becomes grayer, so the absorption of flux absorbs
nearly equally at near-UV and visible wavelengths. 
Consequently, the visible-band flux falls to nearly the same level 
as the UV flux, creating an apparent lag in visible-band light curve. 
Note also that in this interpretation there appears to be no need
to postulate a change in effective temperature of 88\,Her during the 
outburst, as suggested by Doazan et al. (1986) and discussed further
by Smith (2000b). These authors were led to 
this interpretation based on observations showing that the star faded and 
brightened simultaneously over the $\lambda\lambda$1550--5500 region. 
However, we suspect that the redistribution of absorbed light would have
become apparent as a flux {\em brightening} in the near-infrared if the
observations could have been extended to this region as well.

\subsubsection{$\zeta$ Tauri = HD\,37202 }

   $\zeta$ Tau is a well-known B2e star with an extensive flattened 
circumstellar disk which has been imaged using optical interferometric 
techniques (Quirrenbach et al. 1998). Making use of two absorption bumps
at $\lambda$1900 and $\lambda$2400, BBW recently 
fit scattering models to 
UV spectropolarimetric ({\it WUPPE}) data to a temperature of 14,000~K for the 
inner disk. $\zeta$~Tau also exhibits variable line profiles of H$\alpha$ 
and He\,I $\lambda$6678 over a 0.78-day period. Kaye \& Gies 
(1998) initially attributed this periodicity to nonradial pulsations. 
However, Balona \& Kaye (1999) later reinterpreted the profile variations 
in terms of two co-rotating clouds, one at a
low and the other at a high stellar latitude. In the first paper, 
Kaye \& Gies speculated that sharp occasional absorptions in line profiles 
are formed in orbiting cloudlets over the star. 
Thus, the presence of close circumstellar
material (in addition to the star's circumstellar disk) seems to have been
accepted by both groups of authors.

  Figure~\ref{msfig6}a shows that the $\approx$0.8 day period is well defined
in the UVC curve. Using the two pairs of observations indicated in the 
caption, we computed the spectral ratios and computed the mean in spectrum 
in Figure~\ref{msfig6}b. The r.m.s. errors are taken from the two constituent 
ratioed spectra. In computing this mean, we found an offset of 1\% between 
these curves, either because of instrumental reasons
or because the absorptions had actually changed. For presentation in our
figure we have corrected the respective scales of the two ratioed spectra 
by -0.5\% and +0.5\% in order to refer them to a common mean.
Except below $\lambda$1250, the two binned spectra agree quite well.
In approaching the simulation of these observations, we were guided by the 
especially prominent variations of the Al\,III $\lambda$1855--63 doublet
(see Fig.~\ref{msfig6}c),
since the local maximum near $\lambda$1900 in Fig.~\ref{msfig6}b is subtle 
at best. These lines indicate the presence of cool circumstellar matter. 
Using an effective temperature of 21,000~K for $\zeta$\,Tau, (BBW, Harmanec 
2000), a best fit to the line strength variations gave a a cloud 
temperature of 8,000~K ($\pm{2,000}$K), a column density of 
2$\times$10$^{23}$ cm$^{-2}$, and a coverage factor of 20\% to fit the 
ratioed spectrum. Except for the lower column density, this is 
essentially the model that we used in Fig.~\ref{msfig5}b for 88\,Her. We
should add that the spectrum of this model shows a mean absorption 
level for wavelengths above $\lambda$2000 that is $\sim$20\% higher than 
is observed. A similar small disparity occurs for 88\,Her in this narrow
wavelength region. This difference could be due to errors in atomic 
parameters or (in our view, less likely) to an error in the flux linearization 
of {\it NEWSIPS} at these wavelengths. Despite this discreprancy, the 
quality of the fit of the model is good and thus lends support to the Balona 
\& Kaye interpretation of a magnetic co-rotating cloud over this particular 
star. 
 
  Turning our attention to the behavior of prominent spectral lines, we note
the report by Kaye and Gies (1998) that the He\,I $\lambda$6678 line of 
$\zeta$\,Tau can vary in strength by 0.5--1\% of the continuum over several 
hours. Simulations of this line with {\it CIRCUS} show that such variations 
can be reproduced in clouds over a wide range of cloud temperatures. 
To determine a lower temperature limit for the He\,I line variations, 
we ran models to see what temperature ranges could just produce
additional absorptions of 0.5\%. We found the lower limit to the temperature 
to be 11,000~K. Perhaps not coincidentally, this is also the value for 
which $\lambda$6678 becomes optically thick in 
a spherical cloud with the dimensions determined from the 20\% coverage factor.
Because the absorptions from a warm cloud component could be hidden in the 
model shown in Figure~\ref{msfig6}c, the appearance of He\,I line variations
does not contradict the spectrophotometrically derived value of 8,000~K. 
Another indication of the presence of a torus-cloud is variability in the 
Fe\,III ``resonance" line spectrum near $\lambda$1910 (not shown). This
variability is detectable in lines with lower levels up to 6 eV, but not 
much higher, and 
is consistent with absorptions from a warmer gas than suggested from the
Al\,III and iron-curtain lines. 
To simulate these variations, we ran several {\it CIRCUS} models for 
temperatures of 8000--15,000~K with column densities similar to those 
required in our models of the He\,I $\lambda$6678 line. With cloud
temperatures in the 10,000--12,000~K range, our models well  
reproduced the Fe\,III variations. Thus, both the Fe\,III and He\,I 
line variations support a second, warm ($>$10,000~K) cloud component, 
in addition to the cooler one inferred from the Fe-curtain lines. As with
the case of 88\,Her, a warm secondary cloud component could easily be hidden
in a multi-temperature model of the spectrophotometric absorptions.
 
  A few additional comments should be made about the circumstellar body 
indicated by the variations in the curtain and Fe\,III resonance lines. 
The first is that a trade-off between the coverage factor and the column 
density is possible in the fit to the Fig.~\ref{msfig6}b spectrum. We find
that larger coverage factors and higher column densities cause features in
the ratioed spectrum to increase and decrease, respectively. The coverage
factor may be thought of merely as a scaling parameter, so as it decreases
any spectrophotometric details will tend to decrease in visibility. A higher
column density makes the variation smaller too because at this temperature
the effects of foreground absorption are more nearly gray. For these reasons 
we cannot rule out a model with a higher column density of, say, 
5$\times$10$^{23}$ 
cm$^{-2}$ and a coverage factor of 13\%. Our preference for the fit depicted 
in Fig.~\ref{msfig6}b was constrained from the recognition that variations 
of the resonance lines, such as the Al\,III lines depicted in 
Fig.~\ref{msfig6}c, would be difficult to explain with models invoking 
inordinately high column densities and an attendant saturation of the 
strong metallic lines. Both the resonance lines of Al\,III and the Fe\,III 
lines near $\lambda$1910 require model column densities of 10$^{22--23}$ 
cm$^{-2}$, depending on the assumed temperature, to match the observations. 
In addition to a cloud width-length trade-off, we found
that the contrast of the $\lambda$1900 and neighboring bumps in the 
ratioed spectrum can be simulated with multi-temperature cloud models, 
but with a clear loss of uniqueness.

  Thirdly, it is conceivable that the variations of both the He\,I and Fe\,III 
lines could instead arise from a 0.78-day NRP modulation in the 
photosphere. However, to this hypothetical suggestion, one can respond that 
there is additional evidence that the Fe\,III lines in these observations vary 
on long timescales and thus are often unrelated to conditions near the surface.
In investigating the behavior of the Fe\,III lines, we found that
there are considerable differences between the line strengths in 1991 and 
additional observations some four years later. Moreover, by 1995 the star's 
UVC flux had brightened by 10--15\%, whereas the strengths of all 
Fe\,III and Al\,III lines became much weaker. 
The simplest explanation for these changes is that the column length of 
material in front of the star had decreased in this time. We suspect 
that this condition arose from changes in the circumstellar disk rather
than the smaller, co-rotating clouds we have considered in this work. 
Changes in the column density of the disk would not necessarily mean that 
the disk structure had changed during this time. Indeed, it could also 
have changed its orientation. The disk of $\zeta$~Tau is thought to be 
non-axisymmetric in density and to precess around the star with a period 
variously estimated to be 2.3 or 7 years (Delplace 1970, Vakili et al. 1998).  
Hence, long-term changes in the column length of circumstellar material 
could be due to disk precession.

\subsubsection{60 Cygni = HD\,200310 }

   60\,Cyg is a B1Ve star which exhibits light variability with a
2.48 day period (Harmanec et al. 1986). Figure~\ref{msfig7}a depicts the 
UVC light curve for this star taken from {\it IUE SWP}-camera observations 
in July, 1995. This curve appears consistent with this period, but there 
appear to be variations on other timescales as well. Koubsky et al. (2000) 
have come to a similar conclusion from visible-wavelength data. 
These authors also note that the star has a mild helium enhancement, 
which hints already that the star could have a co-rotating magnetosphere.
An inspection of {\it IUE} spectra reveals that the star's spectrum shows
a very strong N\,V doublet as well as variable emission in both this and the 
C\,IV doublets. These are indeed signatures are co-rotating clouds associated
with rotating magnetic B-stars (Smith \& Groote 2001). We next grouped the
{\it IUE} spectra into bright- and faint-star star phases according to this 
period. An example of this comparison is displayed in Figure~\ref{msfig7}b. 
This ratioed-spectrum plot shows a weak but discernible absorption bump at 
$\lambda$1900. This feature, along with the overall, nearly flat, shape of 
the absorption spectrum, can be used to set constrain the temperature of 
the co-rotating cloud inferred from the presence of variable resonance 
lines discussed in the following paragraph. In general, the same 
comments about the geometrical and temperature trade-offs discussed for
$\zeta$\,Tau (Fig.~\ref{msfig6}b) apply to the fit of Fig.~\ref{msfig7}b,
but to a lesser extent. We also exhibit our best fit in Fig.~\ref{msfig7}b, 
based on these trade-offs and a {\it CIRCUS} 
model with T$_{eff}$ = 26,000~K and cloud parameters $T_{cloud}$ = 13,000~K, 
column density = 1$\times$10$^{23}$ cm$^{-2}$, and a coverage factor of 60\%. 

  Figure~\ref{msfig7}c exhibits typical spectral variations of the Al\,III
and a nearby Fe\,II line. Other high excitation lines in the {\it SWP} 
spectra, such as He\,II 
$\lambda$1640, show no detectable variations, indicating that a medium cooler
than the photosphere is responsible for the low-excitation line activity.
Modeling of the variations of these three lines shows probably fortuitously
good agreement with the parameters obtained from panel (b). The average of
the 3 line variations shown in panel (b) is 18 $\pm{3}$\%, and of the modeled 
variations, 21\%. A ``perfect match" to the mean of the variations of these
lines could be made with T$_{cloud}$ = 12,000~K. We estimate the actual 
errors on our fits from the spectral fitting are relatively
large, about $\pm{2500}$~K, because of the large rotational broadening. 
In sum, the evidence from this figure supports an interpretation from the 
strong C\,IV and N\,IV lines of 60\,Cyg (cf. Smith \& Groote 2001) that most 
or all of its UV spectral variations are caused by cloud occultations. 
However, if such clouds exist, the evidence from the light curve and the 
drifts of multiple subfeatures across the optical line profiles (Koubsky et 
al. 2000) suggests an ephemerality reminiscent of the clouds in $\gamma$\,Cas 
(Smith \& Robinson 1999).

\subsection{Pulsation }

  One of the first results to emerge from our study was that a distinct 
spectrophotometric pattern is present in binned ratioed spectra of radially 
pulsating $\beta$~Cephei stars ($\S$\ref{puls}). This result is depicted
for six of these variables with known radial pulsation modes in 
Figure~\ref{msfig8}.  These features are largest in stars having large 
pulsation amplitudes, such as BW\,Vul and $\nu$\,Eri. A simulated ratioed 
spectrum (taken from Fig.~\ref{msfig3}b), is shown for reference.
A comparison of the observations with the simulation shows strong 
similarities in the detailed absorption pattern. This fact suggests that 
they arise from an increased temperature gradient and/or microturbulence. 
Evidently, the ratioing of {\it SWP} spectra provides a new way of determining 
whether radial pulsations are responsible for a star's light and spectral 
variations.

  We also discovered that a large-amplitude, {\em nonradially} pulsating star, 
$\epsilon$~Persei, exhibits a noticeable pulsation signature even 
though all its excited modes are nonradial ($l$ = 3--5; Gies et al. 1999).
The velocity amplitudes of the largest of these modes are very 
large and at times rival the atmospheric sound speed (Smith, Fullerton, 
\& Percy 1987). 
Because the periods are short, the dominant velocities are directed nearly
nearly vertically upwards or downwards at opposing phases. 
These pulsations also cause local temperature and pressure changes, which
can cause a net change in the equivalent width of the line. Evidently, these
effects do not completely cancel when integrated over the visible disk. 
In fact, line strength changes from these pulsations are visible in 
the ratioed spectrum of this star in Fig.~\ref{msfig8} (bottom line). 

  We should point out that in our depiction of the pulsational signatures of
$\beta$~Cep (Fig.\ref{msfig8}) we obtained observations taken 
over just {\em one} pulsation cycle.
Smith \& Groote (2001) used ratioed spectra over different phases of this
12-day {\em rotation} cycle of this same star to demonstrate 
the presence of a co-rotating cloud. In the material used for constructing 
the $\beta$\,Cep curve in our figure,
the cloud absorptions were equally present during both bright- and 
faint-star pulsation phases. Hence, the spectrophotometric ratioing 
technique can be utilized in a several-hour time series 
to detect pulsations in a magnetic star occulted by a cloud.

\subsection{Application to Be Stars}

  The primary goal of our study was to see whether spectrophotometric 
information derived from short-wavelength {\it IUE} observations can be 
used as an tool for candidate variability models operating in Be stars
such as nonradial pulsation and magnetic co-rotating clouds. 
One answer to this question is that the discrimination between the NPR and 
cloud models requires corroborative evidence, such as the selective 
strengthening of low-excitation lines on the timescale of the star's 
rotation period. Table~\ref{tab1} summarizes results for 18 early-to-mid 
Be stars with rapid and arguably periodic UV continuum variations. 
These are stars from a more complete sample 
of early-type Be stars assembled by the author which have been observed 
repeatedly by the {\it IUE} satellite during its lifetime.\footnote{Several 
stars in our original
sample did not exhibit significant UV variations and are not represented 
in the table. These are 59\,Cyg, $\theta$~CBr, X\,Per, 66\,Oph, 48\,Lib,
and $\zeta$\,Oph. The observations of several other demonstrable variable
stars were too widely spaced in time for a reliable analysis. Marginal
results were found for three stars, 28\,CMa, $\kappa$\,CMa, and $\alpha$\,Ara, 
and are discussed following the presentation of Table 1.}
``Group 1" in the table consists of five stars, each of which shows at least 
mild ($\ge$ 8\%) UV variations and for which ratioed spectra exhibit the 
pulsational signatures, discussed in $\S$\ref{puls} These are similar to those
in Figs.~\ref{msfig3} and \ref{msfig8}. 
In Figure~\ref{msfig9}a we exhibit ratioed spectra of $\eta$\,Cen, 
DU\,Eri (HR\,1423) EW\,Lac, $\lambda$\,Eri and which show these same 
signatures. The signatures are 
stable for these stars, being present consistently for different combinations 
of faint- and bright-phase spectra. If $\epsilon$~Per serves as a model,
large-amplitude pulsations can produce strong, local atmospheric gradients 
in temperature and/or velocity which have signatures visible in far-UV spectra. 
From the length of the periods the pulsations must be nonradial pulsations.

  Group~2 in the table consists eight Be stars with periodic UVC variations
that are nearly monotonic with wavelength.
Fig.~\ref{msfig9}b depicts ratioed spectra of 28\,Cyg, $o$\,And, 
$\epsilon$\,Cap, and $\omega$~\,Ori as examples. 
The spectrum for $\alpha$\,Eri, also in this 
category, has already been shown in Fig.~\ref{msfig1}. The ratioed spectra 
of these stars do not show repeatable discrete features, but rather a 
tendency to rise toward short wavelengths. This pattern is consistent 
with the ratio of spectra synthesized from model atmospheres with different
values of T$_{eff}$ and is also in keeping with the 
(small-amplitude) nonradial pulsation paradigm in which spectrum contrasts 
are produced over time as mutually cancelling cool and hot regions travel 
across the disk with phase (Fig.\ref{msfig1}c). Of course, such an explanation 
does not rule out the alternative interpretation of absorbing clouds. However, 
the continua of the clouds' absorption spectrum would necessarily have to 
be optically thick and the coverage factors would have to be nearly 100\%.
In addition, we noticed in our data inspection that the 
resonance lines can have variable amplitudes from one brightness cycle 
to the next. Hence, if the variability of these stars arose from cloud 
absorptions, the clouds would have to be able to change their structures 
over a timescale of about a day. One cannot rule out the presence of large
clouds with such transient properties, but this phenomenology is not 
present in clouds over rotating magnetic B and Bp stars. 
Thus, the spectrophotometric diagnostics favor nonradial pulsation for 
most or all the stars in Group~2 arise from nonradial pulsation as well. 

  The spectrum of $\alpha$\,Eri, previously displayed in Fig.\ref{msfig1}, 
is a case that should be singled out for special discussion. A direct 
comparison of its spectra and its UVC light curve reveals that the photospheric
components of this star's Si\,IV and C\,IV resonance lines are {\em 
anticorrelated} with periodic UVC variations, and they probably have a small 
phase delay. Moreover, time series of optical spectra of this star disclose 
that the He~I $\lambda$5876 and $\lambda$6678 lines can have variations 
which are not morphologically similar (Rivinius 1999). All these behaviors 
are puzzling according to our current understanding of nonradial pulsations. 
Nonetheless, other patterns, such as the variation of the Al\,III doublet, 
can be understood easily within the context of NRP. 
For example, these lines do not have deep cores that would be indicative 
of a cloud or shell at any phase. In addition, the wings of these lines 
strengthen during the ``faint star" phase of a cycle. 
This fact can be explained simply as the result of an ionization shift 
occurring in a high density gas, i.e., at the high temperature extremum 
in a temperature-modulating photosphere. 
Hence this behavior argues for nonradial pulsations in $\alpha$\,Eri.

 Finally, Group~3 of Table~\ref{tab1} consists of three stars with 
cool co-rotating magnetospheres. We have already discussed the arguments
for including 60\,Cyg and $\zeta$\,Tau in this group.
In addition, we include the unusual Be star $\gamma$\,Cas.
We do not discuss this star herein because we have treated it
at length elsewhere (Smith, Robinson, \& Hatzes 1998)
and also because this star's X-ray and UV variability patterns are 
unique.\footnote{However, 
we can add here that variations of the C\,II $\lambda$1335--6 and
Si\,II $\lambda$1200 lines show a clear correlation with the continuum 
light curve, as would be expected for clouds having a cool component.}

  Three other stars, 28~CMa, $\alpha$~Ara and $\kappa$~CMa, which turned 
up in the course of our investigation are not listed in the table. However,
they have redshifted emission in the Si\,IV and C\,IV lines and therefore 
are candidates for having torus-clouds. In the spectrum of 
28 ($\omega$) CMa one finds mild redshifted emission components 
in C\,IV and Si\,IV lines 
which vary on an unknown timescale. These characteristics could be explained 
in principle variable P~Cygni-type wind activity, but probably not for the 
strong N\,V absorptions in the spectrum of this relatively cool (B2.5 IV-V) 
star. Optical work shows that 28\,CMa is also an apparently-single star viewed 
at a nearly pole-on orientation in which the lines exhibit profile variations 
attributed to NRP (Maintz et al. 2000; but see Balona et al. 1999). 
In addition, optical lines formed at low densities and/or temperatures show 
their own special rapid, quasi-periodic behavior, suggesting the presence of a 
rarified structure above the photosphere (Stefl et al. 1998, Stefl et al. 2000). 
The {\it IUE} spectra of the rapidly rotating B2.5Vne
star $\alpha$\,Ara likewise exhibit redshifted emissions in the C\,IV and 
Si\,IV lines that are variable even over 2 hours. This star
has neither a variable radial velocity history (Thackeray 1966) nor a high 
X-ray flux (Cohen, Cassinelli, \& MacFarlane 1997) that could be indicate the
presence of hot gas streams in an interactive binary. 
A third conceivable cloud-bearing 
candidate, $\kappa$\,CMa (B1.5IVne), was observed with the {\it IUE SWP} only 
three times, but its C\,IV and Si\,IV lines once again show weak redshifted 
activity and possible emission, though on an indeterminate timescale. N\,V 
absorptions are also visible. Balona (1990) has noted rapid photometric 
variations consistent with the rotational period during an outburst of 
this star.

\subsection{Conclusions} 

  This study has demonstrated that high-dispersion {\it IUE} archival
spectra of variable B stars can be manipulated to bring out qualitatively 
new information about the mechanisms responsible for these variations.
Bjorkman, Bjorkman, \& Wood (2000) have recently used UV spectrophotometry 
to determine physical attributes of circumstellar material around Be stars. 
In contrast to our technique, their study exploited the polarization 
signatures of asymmetrical {\em large-scale} circumstellar structures 
from single observations. Although our study has been limited to {\it 
IUE SWP} data, it also could be extended in principle to long-wavelength 
camera data of large-amplitude late-B, A-, and F-type variables. 

  In $\S$\ref{analy} we demonstrated how quantitative information can be 
gleened about gas ejected by hot stars. Our examples
included Be stars which are surrounded by a compressed wind-wake, eject shells, 
have co-rotating clouds, or undergo large-amplitude pulsation. Indeed, 
our binned spectrum-ratio results demonstrate that quasi-monochromatic
spectrophotometic changes of a Be star do not necessary imply the 
variation of its photospheric temperature from pulsation (cf. Peters 1998b),
although this may well be the rule. The technique does not work well for 
determining parameters of inhomogeneous clouds or even of cool, dense 
clouds with small projected areas. Even so, we have seen 
that the presence of a warm cloud component may be inferred from a spectral 
analysis of specific lines in the same observations. Because residual 
errors in flux calibrations are still present, {\it IUE} data are limited 
in how they can be utilized over extended time periods. Still, they are useful 
in the some cases. For 88\,Her in particular the progressive change in
line strengths during an outburst shows that the ``phase lag" of a
dip in the {\em visible-band} light curve can be understood by the cooling 
of an expanding shell. Thus, the technique can compliment post-outburst 
observations of high-level Balmer lines and infrared emission studies in Be 
stars. It can be particularly helpful in quantifying 
changes of shell parameters with distance from the star.

  Although one goal of this study was to diagnose the physical cause(s)
of light and absorption line variability in classical Be stars, our success 
in using UV spectrophotometry for this purpose must be said to be mixed: 
potential ambiguities could arise between dense clouds and weak NRPs in some 
cases. According to our ``Group~3" list, Be stars with candidate co-rotating 
clouds actually show identifiable 
spectrophotometric signatures in only three of five cases. 
Even leaving aside such unique stars as $\gamma$~Cas, a good case can be 
made for the existence of magnetic co-rotating clouds in only a minority of 
Be stars. Aside from the parenthesized entries our sample is relatively 
unbiased of variable early-to-mid B stars (as far as we know), so one may 
take the implied fraction of them having clouds to be about 1/8--1/4.
These rough statistics suggest a magnetic cause (or trigger) for
the Be phenomenon in at least a minority of classical Be stars.
The existence of co-rotating clouds in particular seems to be a 
by-product of {\it dipolar} fields (Shore \& Brown 1990)
However, evidence exists also for {\em localized}, transient magnetic 
activity in other Be stars. Such evidence includes rapid line 
profile variability (e.g., ``dimples"; Smith \& Polidan 1993), and 
in the case of $\lambda$\,Eri, the appearance of an X-ray flare. 
Thus, it is possible that
localized fields can play a role on the surfaces of other Be stars. 
Unfortunately, hypothetical prominence-type clouds associated with 
small-scale magnetic loops on Be stars (e.g., Peters 1998b)
could not be detected with ratioed {\it IUE } spectra.
Thus, we look to future optical studies of cloud-candidate Be stars 
to confirm the results of this work. 
Evidence of transits of low-density and temperature clouds might be found 
in high-quality time-series spectra, both in the appearance high-level 
Balmer lines and as occasional brief events observed in many lines. 

  We wish to acknowledge our appreciation to Dr. Ivan Hubeny for making
his {\it SYNSPEC} and {\it CIRCUS} codes available to us. We also thank 
Dr. Dietrich Baade who made several comments that led to substantial 
improvements in this paper and also several helpful comments by the referee.
This work was conducted in part under NASA NAG 5-8793.

%


\setlength{\textwidth}{183mm}
\setlength{\evensidemargin}{22mm}

\begin{table}[ht!]
\begin{center}
\caption{\label{tab1} Summary of Spectrophotometric Characteristics
for Be Stars }
\centerline{~}

\begin{tabular}{cccc} \hline \hline
Group 1 & (Pulsation) &    &   \\
    &  Sp.   & Period (d)   & Vsin\,i   \\ 
\hline
DU Eri  &  B2V:ne & 1.2\tablenotemark{1}  &  340  \\
$\lambda$\,Eri  & B2IVne  & 0.701\tablenotemark{2}  &  310  \\
$\eta$\,Cen  &   B1Vne  & 0.642\tablenotemark{3}  & 350  \\ 
2\,Vul     & B0.5IV   &  0.35:\tablenotemark{1}    &  332   \\     
EW\,Lac  &  B3IVpe  &  0.722\tablenotemark{4}  &  340  \\       
\hline\hline
   Group 2    &    &  &  \\
\hline
$\alpha$\,Eri  & B3pe    & 1.26\tablenotemark{5}   &  220  \\  
$\psi$\,Per    & B5Ve    & 1.04\tablenotemark{6}  & 369  \\   
$\omega$\,Ori  & B3IIIe  & 1.9\tablenotemark{2}  & 160 \\   
PP\,Car      & B4Vne  & 0.8:\tablenotemark{7}  & 303   \\ 
28\,Cyg      & B2.5Ve & 0.646\tablenotemark{8} & 310    \\
$\epsilon$\,Cap   & B3V:p   & 0.98\tablenotemark{1}  & 293  \\
120\,Tau     &  B1.5IVe  &  0.5:\tablenotemark{1}  & 271  \\
$o$\,And     &  B6IIIpe  &  1.58\tablenotemark{8}  & 330     \\
\hline\hline
Group 3    & (Cloud-like)   &     &     \\  \hline
$\zeta$\,Tau  &  B4IIIe  & 0.78\tablenotemark{10}  & 320 \\
60\,Cyg     &  B1Ve    & 2.48\tablenotemark{11}     & 220  \\
$\gamma$\,Cas & B0.5IVe     & 1.12:\tablenotemark{12} & 220     \\
\hline
\end{tabular}

\end{center}
	\end{table}

\newpage 
$^{1}$Peters \& Gies 2000,
$^{2}$Balona et al. 1987,
$^{3}$Balona 1999,
$^{4}$Floquet et al. 2000,
$^{5}$Balona et al. 1987,
$^{6}$Peters 1994,
$^{7}$this paper,
$^{8}$Tubbesing et al. 2000,
$^{9}$Sareyan et al. 1998,
$^{10}$Kaye \& Gies 1998,
$^{11}$Harmanec et al. 1986,
$^{12}$Smith et al. 1998.  

\newpage

\begin{center}
{\bf Figure Captions}
\end{center}

\begin{description}

\item [Figure~\ref{msfig1}: ]    
The binned ratioed spectrum of two {\it SWP}-camera echellograms
of $\alpha$~Eri (SWP~55889 and 55873) processed through two different
spectral extraction algorithms, {\it NEWSIPS} and {\it IUESIPS}. The dotted
line shows the effect of an assumed 2\% error in background level on
this ratio. Note its rapid increase below $\lambda$1250.

\item [Figure~\ref{msfig2}: ]   
Computed binned absorption spectra of a homogeneous cloud for the variety
of temperatures indicated, a column density of 1$\times$10$^{23}$, and a
microturbulence of 20 km~s$^{-1}$.
The mean photospheric temperature for these examples is 22,000~K.

\item [Figure~\ref{msfig3}: ]   
{\bf (a)}~
Ratio of synthesized photospheric spectra over the wavelength range of the
{\it IUE} {\it SWP} camera for Kurucz models having log~g = 4 and T$_{eff}$ = 
26,000~K and 22,000~K. 
{\bf (b)}~
Run of temperature with the atmospheric parameter ``rhox" for a 22,000~K, log~g 
Kurucz model (dashed line), as altered artificially to mimic a low and high
thermal gradient (solid lines).
{\bf (c)}~
The far-ultraviolet ratioed spectrum resulting from the two altered 
temperature gradient distributions shown in panel {\bf (b)}.~

\item [Figure~\ref{msfig4}: ]   
{\bf (a)}~
UV continuum light curve of Vela~X-1, with $\phi$ = 0.0 reckoned from
eclipse of the neutron star {\em behind} the B star 
at HJD 2444279.0466 (van Kerkwijk et al. 1995)
for all available high-dispersion {\it SWP} observations over epochs 
1978--93.
{\bf (b)}~
Model and {\it IUE} spectrum ratio are exhibited as dashed and solid 
lines, respectively for orbital phases
0.4--0.46  (SWP~01442, 18823, 22324, 32961, 46167, \& 49202) divided by
observations taken near phase 0.3 (SWP19061-2, 22309 \& 46151; 
see arrow in panel (a)). 
{\bf (c)}~
Model and observed spectrum ratio are shown as dashed and solid lines,
respectively, for phases 0.5--0.6 
(SWP~03510, 03649, 138958, 19012, 25881-2, 32961, 32967, 49086), again 
divided by observations taken near phase 0.3 (as in panel b). Note the high
temperature and column density and low coverage for the best-fit cloud 
model. This suggests an hot circumstellar structure which partially obscures
the B supergiant primary just after transit of the orbiting X-ray pulsar.

\item [Figure~\ref{msfig5}: ]   
{\bf (a)}~ 
The UVC light curve for 88\,Her in 1981--1992. 
{\bf (b)}~ 
Spectral ratios of 1986 spectra (SWP\,27266, 27672, \& 28334) relative 
to $\sim$1982 spectra (SWP\,14701, 15012, \& 16287).
{\bf (c)}~ 
Comparison of spectra in the $\lambda$1910 region. 
The dashed line represents the spectrum from SWP\,31189; solid
line from SWP~38786 \& 44980. Note that the strengths of the high-excitation 
Fe\,III ($\approx$10\,eV) lines (asterisked symbols) are highly variable 
compared to the two Fe\,III 3.7 eV lines (labeled). The X-symbol denotes pixels
excluded because of an instrumental reseau.

\item [Figure~\ref{msfig6}: ]   
{\bf (a)}~ 
The UVC light curve for $\zeta$\,Tau. Observations 
indicated by arrows are SWP\,42644, 42650, 42658, \& 42664. 
{\bf (b)}~ 
Solid line shows spectral ratio taken from the binned fluxes of 
SWP\,42644 \& 42658 divided by SWP\,42650 \& 42664 (arrows in {\bf a}); 
dashed line is from a {\it CIRCUS} model with the indicated parameters.
{\bf (c)}~ 
Comparison of spectra in the $\lambda$1860 (Al\,III line) region. 
The top line represents the mean spectrum from all 1991 observations. The
bottom line is the r.m.s. spectrum, with line excitations indicated in eV.

\item [Figure~\ref{msfig7}: ]   
{\bf (a)}~ 
The UVC light curve for 60\,Cyg in late July, 1995.  
{\bf (b)}~ 
Solid line shows spectral ratio for 60\,Cyg taken from the binned 
fluxes of SWP\,55536 divided by SWP\,55608 (arrows in {\bf a}); 
dashed line is from a {\it CIRCUS} model with the indicated parameters.
{\bf (c)}~ 
Comparison of the same two spectra in the $\lambda$1855-63 (Al\,III line) 
region.
The bottom line is the r.m.s. spectrum, with line excitations indicated in eV.
This panel shows that even that variations of low-excitation lines are
visible even in this broad-lined star.

\item [Figure~\ref{msfig8}: ]   
Montage of ratioed spectra of {\em known} $\beta$~Cep stars with primarily 
radially pulsating modes 
and one large-amplitude nonradially pulsating B star, $\epsilon$~Per.
Offsets in a few cases are noted for clarity. The dot-dashed line depicts the 
simulated ratioed spectrum resulting from division of spectra with different 
T($\rho$\,x) distributions, as shown in Fig.\ref{msfig3}c. The dashed line 
represents the ratioed spectrum of the large-amplitude, nonradially pulsating 
star, $\epsilon$~Per. These ratioed spectra exhibit pulsation signatures
discussed in the text which are used in Table~1 to differentiate between 
Groups 1 \& 2. The ratioed spectra shown (maximum-flux phase divided 
by minimum-flux) are given by {\it SWP} sequence number as follows: BW\,Vul, 
05590-1, 05597-8 vs. 05594-5; $\nu$\,Eri, 04355 vs. 04351-2; 12\,Lac, 55699 
vs. 55710; $\alpha$\,Lup, 04572-4 vs. 04564-6; $\delta$\,Cet, 04487-8 vs. 
04490-1; $\beta$\,Cep, 04596-7 vs. 04598-9;
and $\epsilon$\,Per, 56521-2, 56577-8 vs. 56524-6.

\item [Figure~\ref{msfig9}: ]   
{\it a}~
Montage of ratioed spectra for {\em presumed} pulsating stars. The upper two
spectra have the $\lambda\lambda$1240--1380 ``window" signature of strong
(indicated by bar with downward errors and spike at $\lambda$1300) and a
plateau at $\lambda\lambda$1400--1540 (bar with upward arrows), as expected
from strong NRP or radial pulsations, according to Fig.~8. 
{\it b}~
Display of ratioed spectra with a nearly a monotonic form with wavelength, 
as expected from simple models of small-amplitude NRP (Group~2). {\it IUE}
sequence numbers for ratioed spectra for bright and faint-star phases for
the two panels are as follows:
$\eta$\,Cen, 41218, 41230, 41254, 41257 vs. 41213, 41234, 41237, 41248, 41260; 
$\lambda$\,Eri, 32228, 32246, 32262 vs. 32238, 32254;
EW\,Lac, 48556, 48558, 48560 vs. 48546, 48548, 48564, 48566, 48568; 
$\omega$\,Ori, 56751--55 vs. 56739--43,
DU\,Eri, 55888 vs. 55882, 55892; $o$\,And, 31479, 31481 vs. 31489, 31491; 
$\epsilon$\,Cap, 34367, 34369, 34386, 34388, 34390, 34392 vs. 34378, 34380;
and 28\,Cyg, 37132, 37134 vs. 37100, 37102. 

\end{description}   

\begin{figure}
\vspace*{.1in}
\caption{msfig1}
\label{msfig1}
\end{figure}

\begin{figure}
\vspace*{.1in}
\caption{msfig2}  
\label{msfig2}
\end{figure}

\begin{figure}
\vspace*{.1in}
\caption{msfig3}  
\label{msfig3}
\end{figure}

\begin{figure}
\vspace*{.1in}
\caption{msfig4}  
\label{msfig4}
\end{figure}

\begin{figure}
\vspace*{.1in}
\caption{msfig5}  
\label{msfig5}
\end{figure}

\begin{figure}
\vspace*{.1in}
\caption{msfig6}  
\label{msfig6}
\end{figure}
  
\begin{figure}
\vspace*{.1in}
\caption{msfig7}  
\label{msfig7}
\end{figure}

\begin{figure}
\vspace*{.1in}
\caption{msfig8}  
\label{msfig8}
\end{figure}

\begin{figure}
\vspace*{.1in}
\caption{msfig9}  
\label{msfig9}
\end{figure}


\newpage
                                                 
\begin{references}


\reference{} Baade, D. 2000, The Be Phenomenon in Early-Type Stars, 
op. cit., p. 178

\reference{} Baade, D. \& Balona, L. A., 1994 Pulsation, Rotation, \& 
Mass Loss in Early-Type stars, eds. L. Balona, H. Henrichs, \& J. 
Le\,Contel (Dordrecht: Kluwer), p. 311

\reference{} Balona, L. A. 1990, MNRAS, 245, 92

\reference{} Balona, L. A., 1995, MNRAS, 277, 1547

\reference{} Balona, L. A. 1999, MNRAS, 306, 407

\reference{} Balona, L. A. 2000, The Be Phenomenon in Early-Type Stars,
{\it op. cit.}, p. 1

\reference{} Balona, L. A., Engelbrecht, C. A., \& Marang, F. 1987,
MNRAS, 227, 123

\reference{} Balona, L. A., Kaye, A. B. 1999, \apj, 521, 407

\reference{} Balona, L. A., Marang, F. et al. 1987, 
A. \& A. S., 71, 24

\reference{} Barylak, M., \& Doazan, V. 1986, A. \& A., 159, 65

\reference{} Bjorkman, K. S., Bjorkman, J. E., \& Wood, K. 2000, The Be
Phenomenon in Early-Type Stars, ed. M. A. Smith, H. F. Henrichs, \& J.
Fabregat, ASP Conf. Ser. 214, p. 603 (BBW)

\reference{} Blondin, J. M., Kallman, T. R., Fryxell, B. A., \& Taam, 
R. E. 1990, \apj, 356, 591


\reference{} Burger, M., de Jager, C., \& van den Oord, G. H. J. 1982,
A. \& A., 109, 289

\reference{} Campos, A. J., \& Smith, M. A. 1981, \apj, 238, 250

\reference{} Cohen, D. H., Cassinelli, J. P., \& MacFarlane, J. J. 1997,
\apj, 487, 867

\reference{} Cugier, H., Daszynska, J., \& Polubek, G. 1998, Rotation, 
Pulsation, \& Mass Loss in Early Type Stars, {\it op. cit.}, p. 17 

\reference{} Danezis, E., \& Theodossiou, E. 1988, Ast. \& Astrophys. 
Suppl., 72, 497
 
\reference{} Danezis, E., \& Theodossiou, E. 1990, Ap. \& Sp. Sci., 72, 497

\reference{} Delplace, A. M. 1997, A. \& A., 7, 459 

\reference{} Doazan, V., Harmanec, P., et al. 1982, Astr. Ap. Sup., 50, 481

\reference{} Doazan, V., Thomas, R. N., \& Barylak, M. 1986, A. \& A., 
159, 75

\reference{} Floquet, M., Hubert, A. M., et al. 1998, A. \& A., 335, 565


\reference{} Furenlid, I., Meylan, T., Young, A., Haag, C., Crinklaw, G. 1987,
\apj, 319, 264

\reference{} Garhart, M. P., Smith, M. A., Levay, K. L, Thompson, R. W., 
\& Turnrose, B. E. 1997, NEWSIPS Manual, IUE NASA Newsletter No. 57

\reference{} Gies, D. R., Kambe, E. et al. 1999, \apj, 525, 420

\reference{} Harmanec, P. 2000, The Be Phenomenon in Early-Type Stars, 
{\it op cit.}, p. 13

\reference{} Harmanec, P., Horn, J. et al. 1978, Bull. Astron. Inst. Czech.,
31, 144

\reference{} Harmanec, P., Horn, J., et al. 1986, IBVS No. 2912

\reference{} Harmanec, P., Koubsky, P, \& Krpata, J. 1974, A. \& A., 33, 117

\reference{} Hubeny, I. 1996, ``Complete Guide to Generate Stellar Spectra
 under the Influence of Absorbing and/or Emitting Bodies," priv. comm.

\reference{} Hubeny, I. \& Heap, S. R. 1996, \apj, 470, 1144

\reference{} Hubeny, I., Lanz, T., and Jeffery, S. 1994, Newsletter on
Analysis of Astronomical Spectra, 20, 30


\reference{} Kaper, L., Hammerschlag-Hensberge, G., \& van Loon, J. 1993, 
A. \& A., 279, 485

\reference{} Kaper, L., Hammerschlag-Hensberge, G., \& Zuiderwijk, E. J. 
1994, A. \& A., 289, 846

\reference{} Kaye, A. B., \& Gies, D. R. 1998, \apj, 482, 1028

\reference{} Koubsky, P., Harmanec, H. et al. 2000, The Be Phenomenon 
in Early-Type Stars, {\it op cit.}, p. 280

\reference{} Kurucz, R.L. 1990, Trans. IAU, 20B, 169

\reference{} Kurucz, R. L. 1993, ``ATLAS9 Stellar Atmospheres and 2
km~s$^{-1}$ Grids," Kurucz CD-ROM \#13

\reference{} Leister, N. V., Janot-Pacheco, E., Leyton, Z. J., Hubert, A.
M., \& Floquet, M. 2000, The Be Phenomenon in Early-Type Stars, op. cit., 
p. 272

\reference{} Maintz, M., Rivinius, T., Tubbesing, S., Wolf, B., Stefl,
S., \& Baade, D. 2000, The Be Phenomenon in Early-Type Stars, op. cit., p. 244

\reference{} Marlborough, J. M. 2000, The Be Phenomenon in Early-Type Stars, 
{\it op. cit.}, p. 743

\reference{}Mathias, P., Gillet, D., \& Crowe, R. 1992, A. \& A., 257, 681

\reference{} Pavlovski, K., Harmanec, P. et al. 1997, Astr. Ap. Sup., 125, 75


\reference{} Peters, G. J. 1991, Rapid Variability of OB-Stars, ed. D.
Baade, ESA Conf. Proc. No. 36, p. 171

\reference{} Peters, G. J. 1994, Pulsation, Rotation, \& Mass Loss in 
Early-Type Stars, ed. L. Balona, H. Henrichs, J. Le Contel (Dordrecht:
Kluwer), p 284

\reference{} Peters, G. J. 1998a, \apj, 502, L59

\reference{} Peters, G. J. 1998b, Cyclical Variability of Stellar Winds,
eds. L. Kaper \& A. W. Fullerton (Springer: Berlin), p. 127

\reference{} Peters, G. J., \& Gies, D. R. 2000, The Be Phenomenon in 
Early-Type Stars, op. cit., p. 375

\reference{} Peters, G. J., \& Polidan, R. S. 1984, \apj, 283, 745

\reference{} Rivinius, T. 1999, priv. communication.

\reference{} Quirrenbach, A., Bjorkman, K. S. et al. 1998, \apj, 479, 477

\reference{} Sareyan, J. P., Gonzalez-Bedolla, S. et al. 1998, A. \& A., 
332, 155

\reference{} Shore, S. N., \& Brown, D. N. 1990, \apj, 365, 665 
 

\reference{} Smith, M. A. 1983, \apj, 265, 338

\reference{} Smith, M. A. 1999a, \pasp, 111, 722

\reference{} Smith, M. A. 1999b, \pasp, 111, 1472

\reference{} Smith, M. A. 2000a, The Be Phenomenon in Early-Type Stars, 
{\it op cit.}, p. 216

\reference{} Smith, M. A. 2000b, The Be Phenomenon in Early-Type Stars, 
{\it op. cit.}, p. 292

\reference{} Smith, M. A., Fullerton, A. W., \& Percy, J. R. 1987, \apj,
320, 768

\reference{} Smith, M. A., \& Groote, D. 2001, A. \& A., in press

\reference{} Smith, M. A., \& Polidan, R. P. 1993, \apj, 367, 302

\reference{} Smith, M. A. \& Robinson, R. D. 1999, \apj, 517, 866

\reference{} Smith, M. A., Robinson, R. D., \& Hatzes, A. P.
1998, \apj, 507, 945

\reference{} Smith, M. A., \& Groote, D. 2001, A. \& A., in press

\reference{} Stamford, P. A., \& Watson, R. D. 1981, Proc. Astr. Soc. Aust.,
4, 210


\reference{} Stefl, S., Baade, D., et al. 1998, A Half Century of Stellar
Pulsations, ed. P. Bradley \& J. Guzik, ASP Conf. Ser. 135, 348

\reference{} Stefl, S., Budovicova, A., et al. 2000, The Be Phenomenon in
Early-Type Stars, {\it op. cit.}, p. 240

\reference{} Thackeray, A. D., 1966, MmRAS, 70, 33

\reference{} Tubbesing, S., Rivinius, T., \& Wolf, B., 2000, The Be
Phenomenon in Early-Type Stars, op. cit., p. 232

\reference{} Vakili, F., Mourard, D., et al. 1998, A. \& A., 335, 261

\reference{} van Kerkwijk, M. H., van Paradijs, J., Zuiderwijk, E. J.,
Hammerschlag-Hensberge, G., Kaper, L., \& Sterken, C. 1995, A. \& A., 303, 483


%
\end{references}
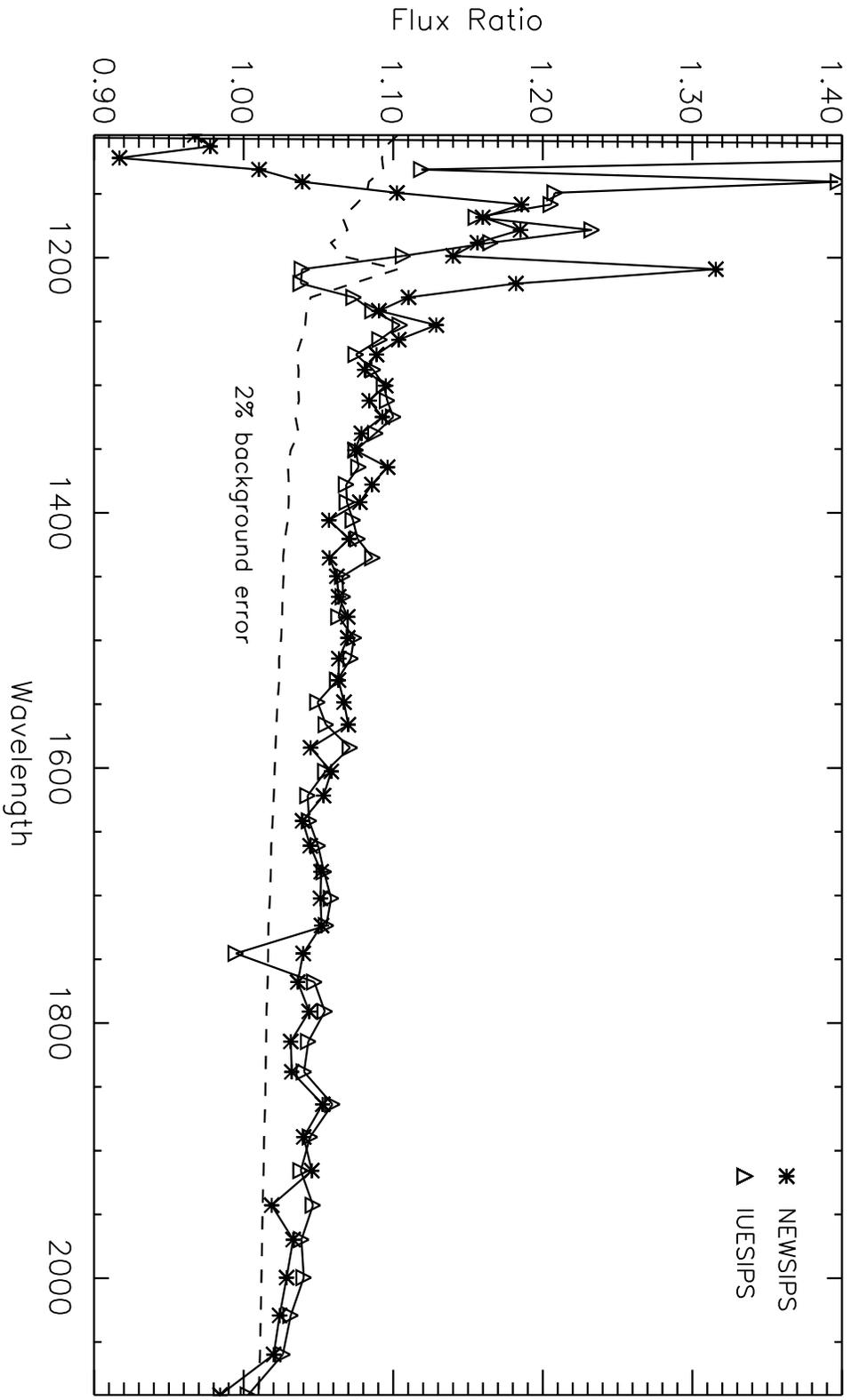
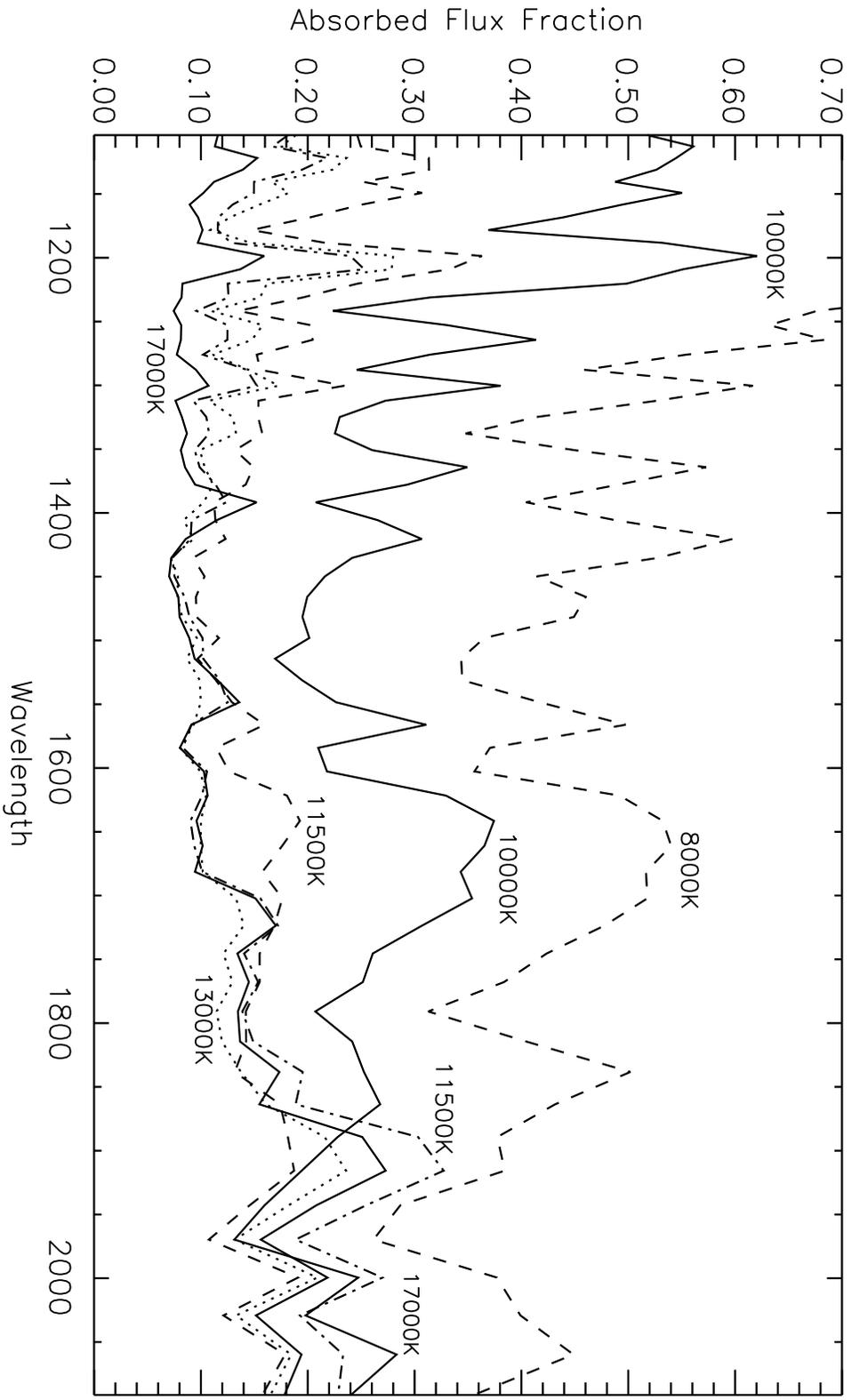
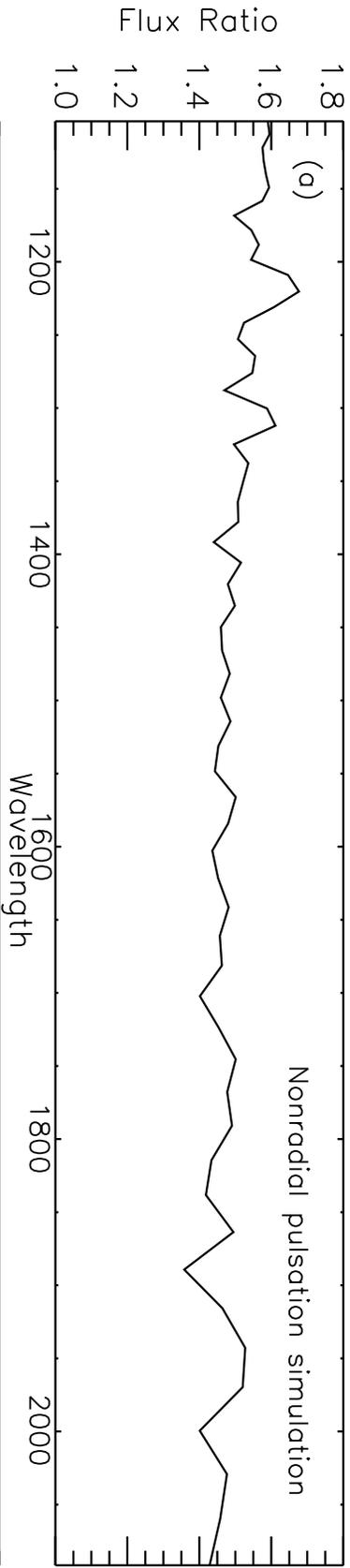
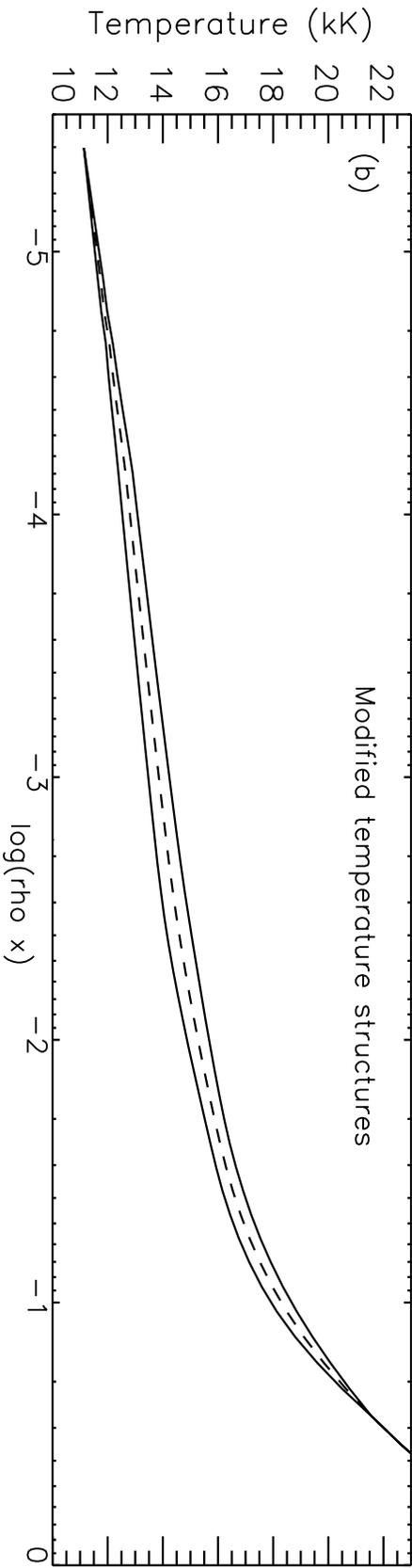
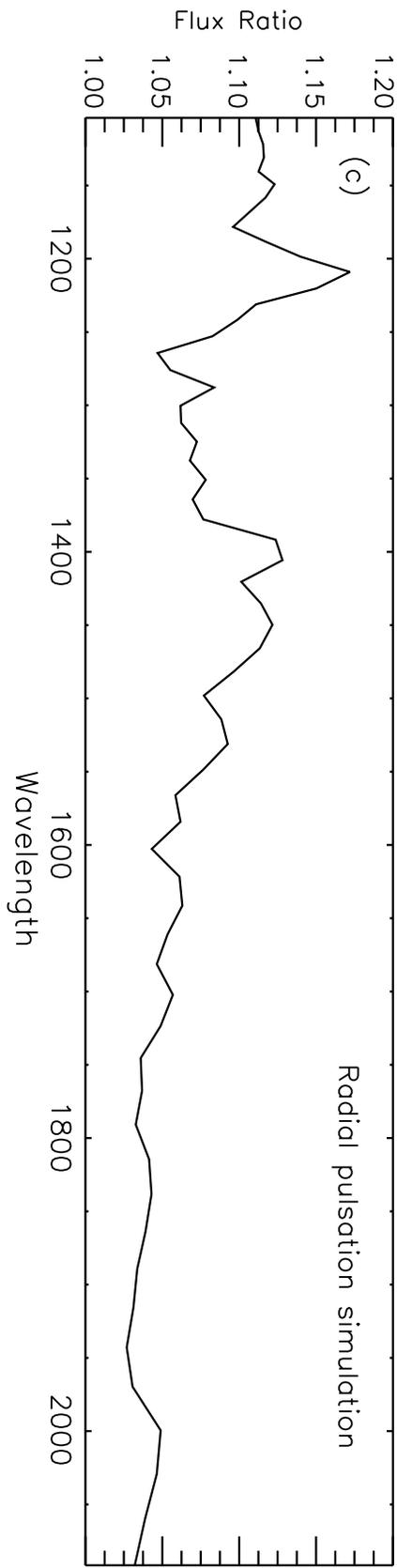
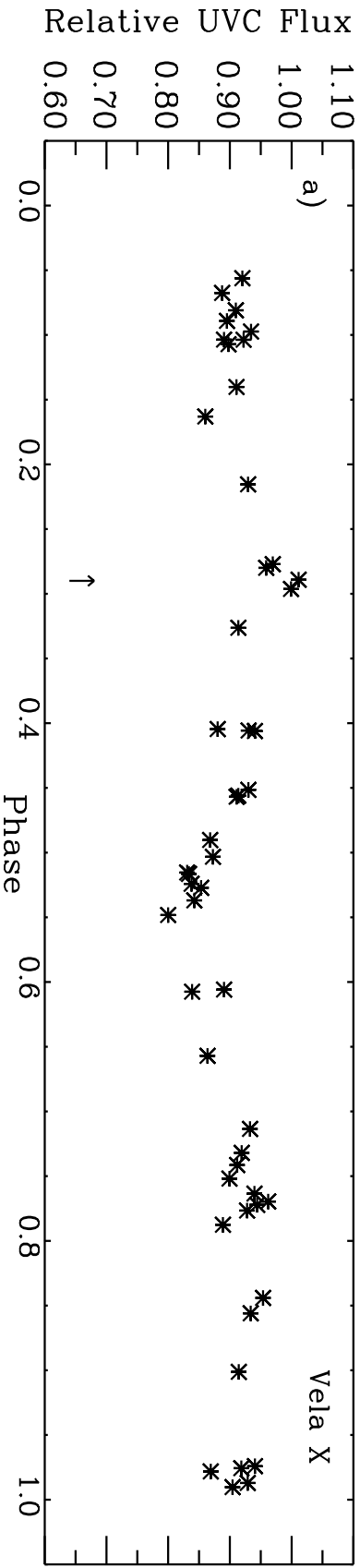
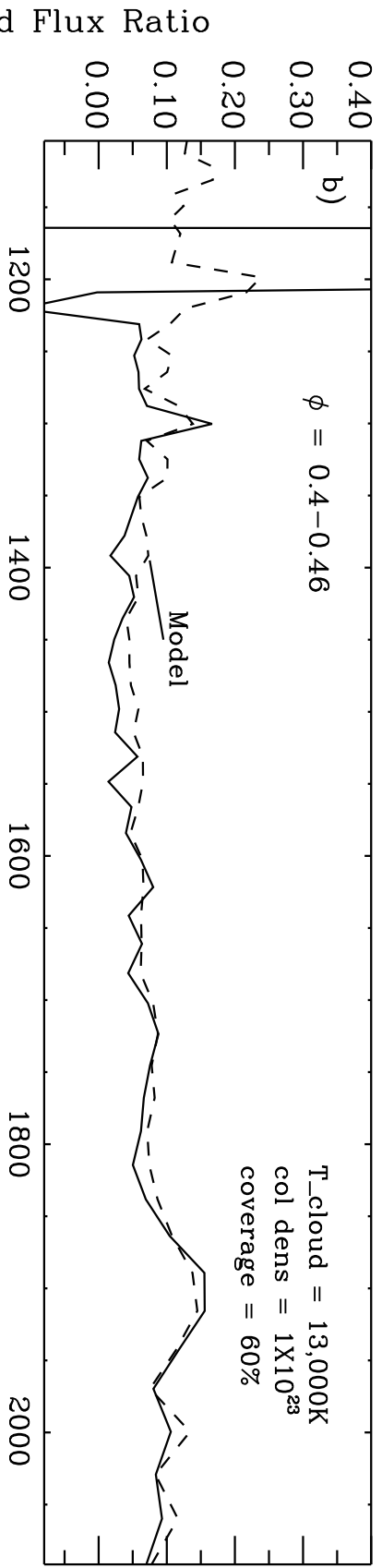
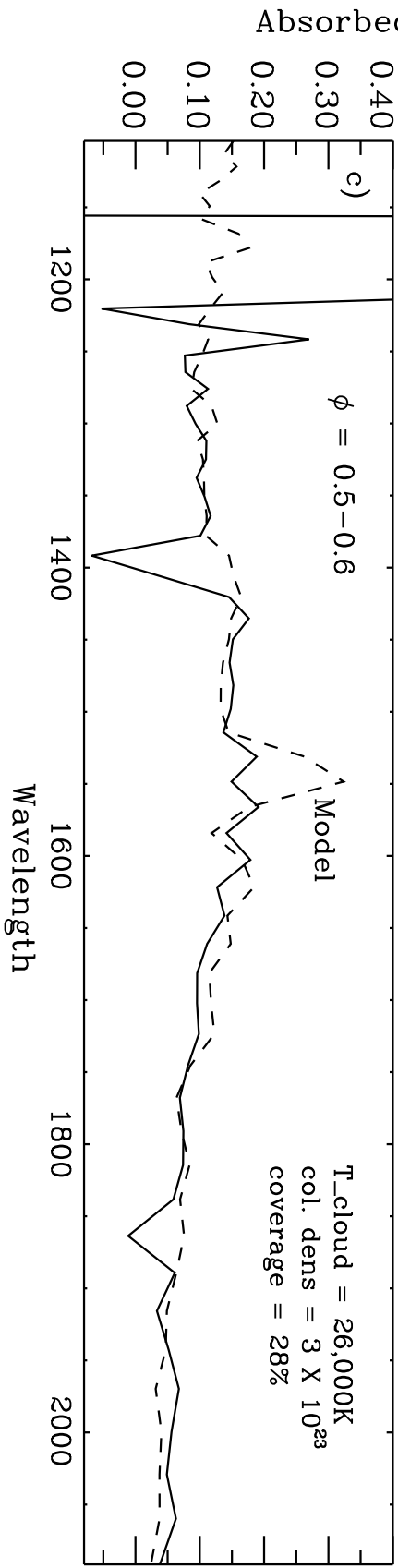
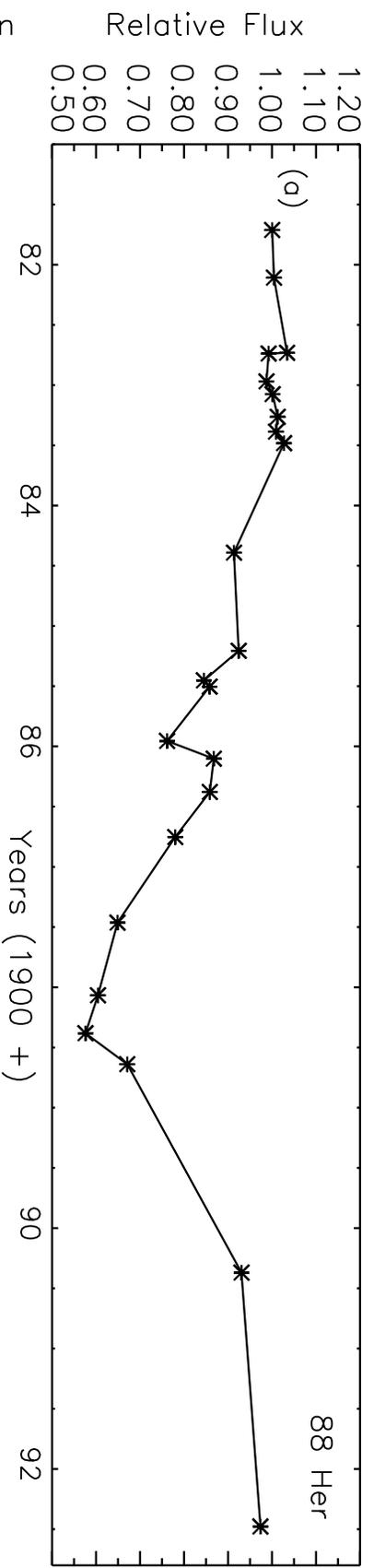
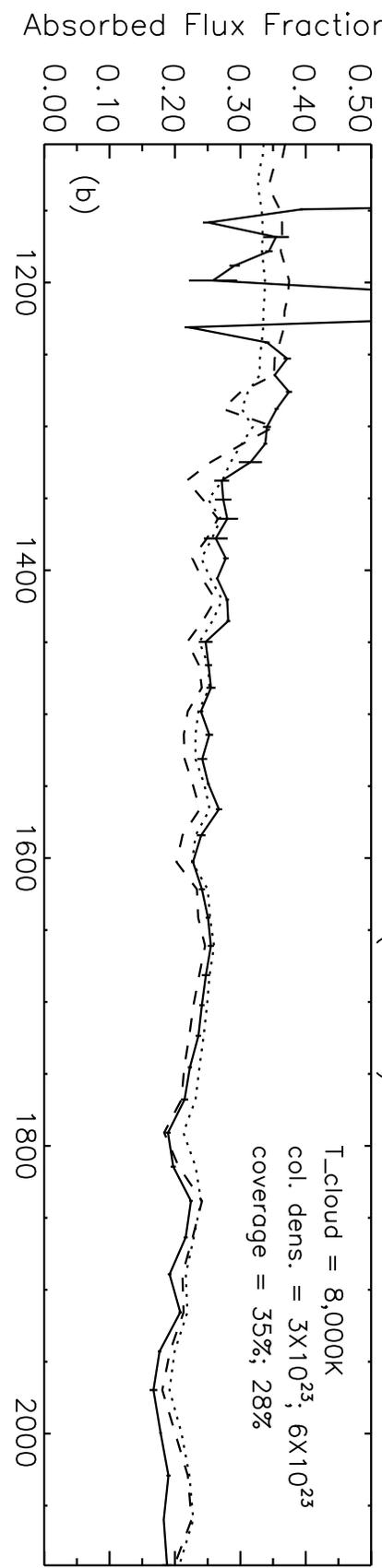
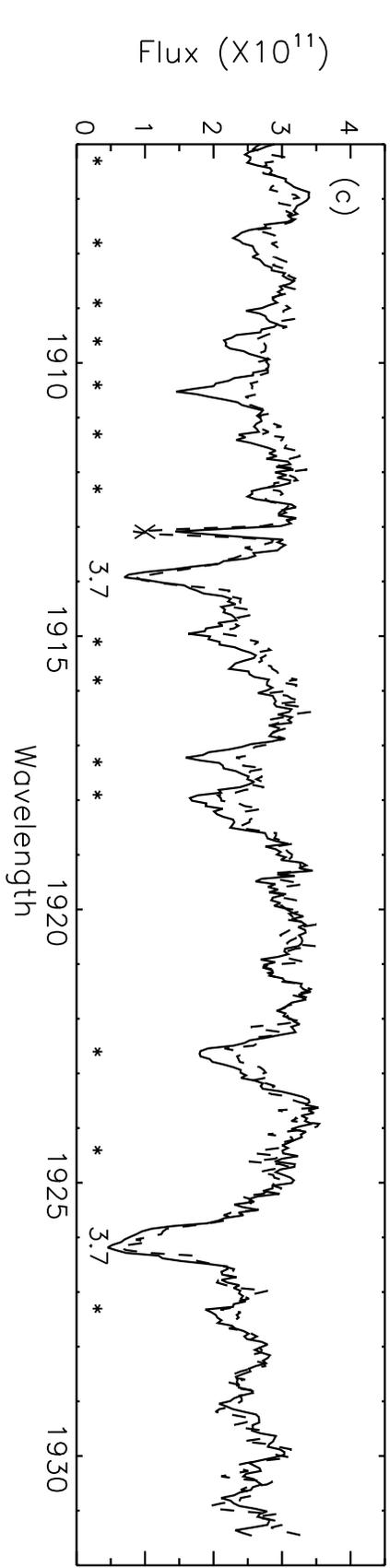
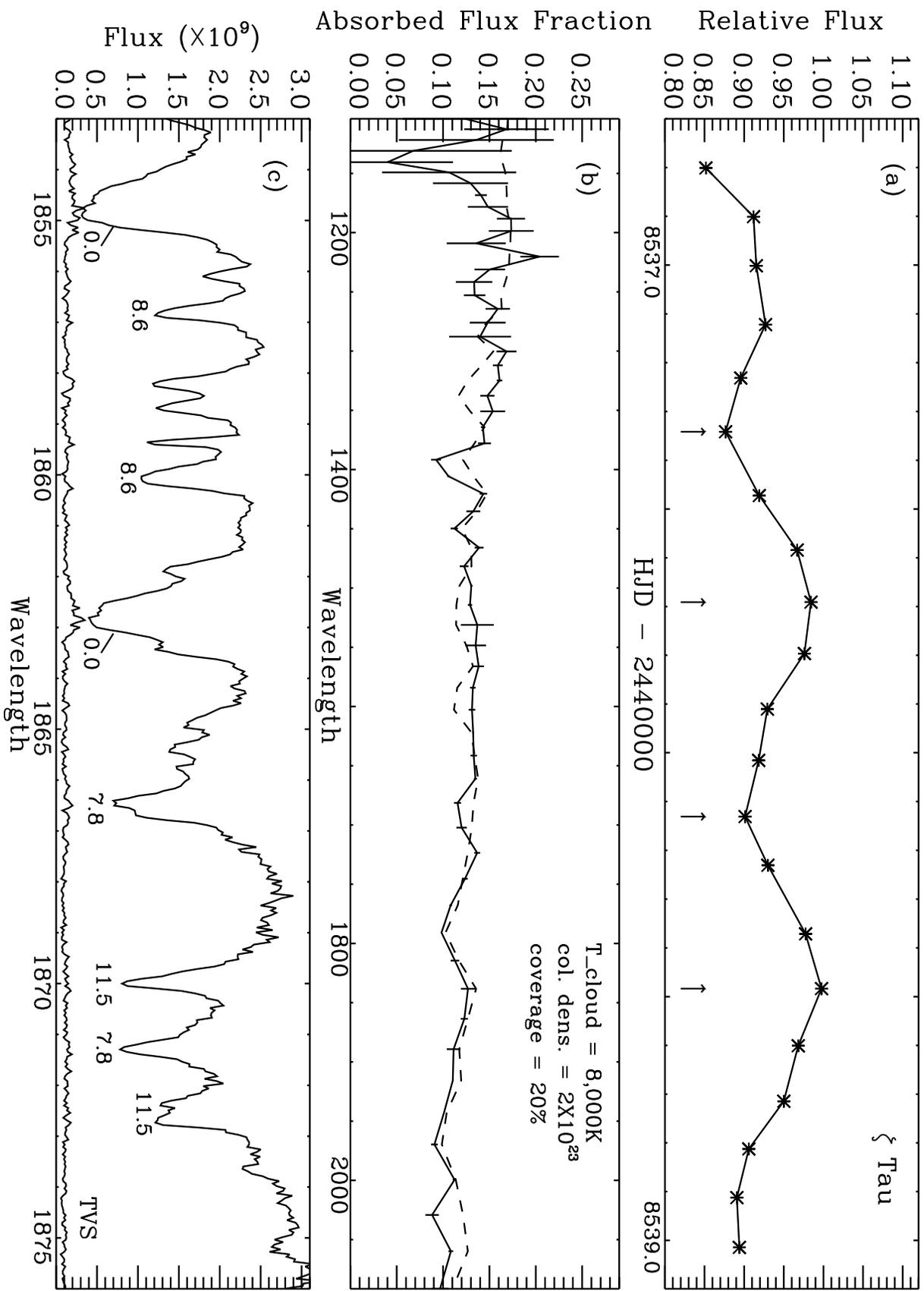
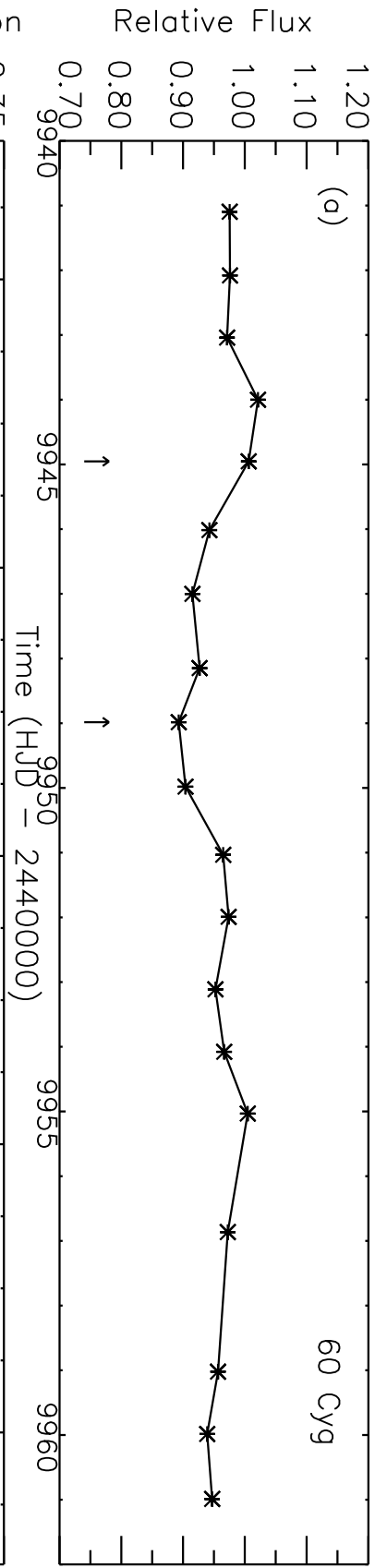
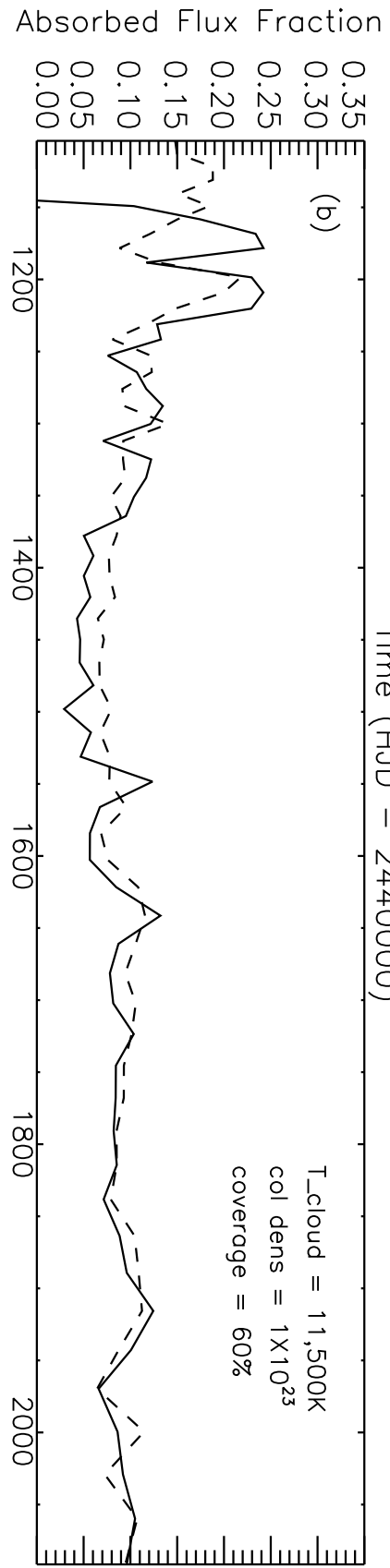
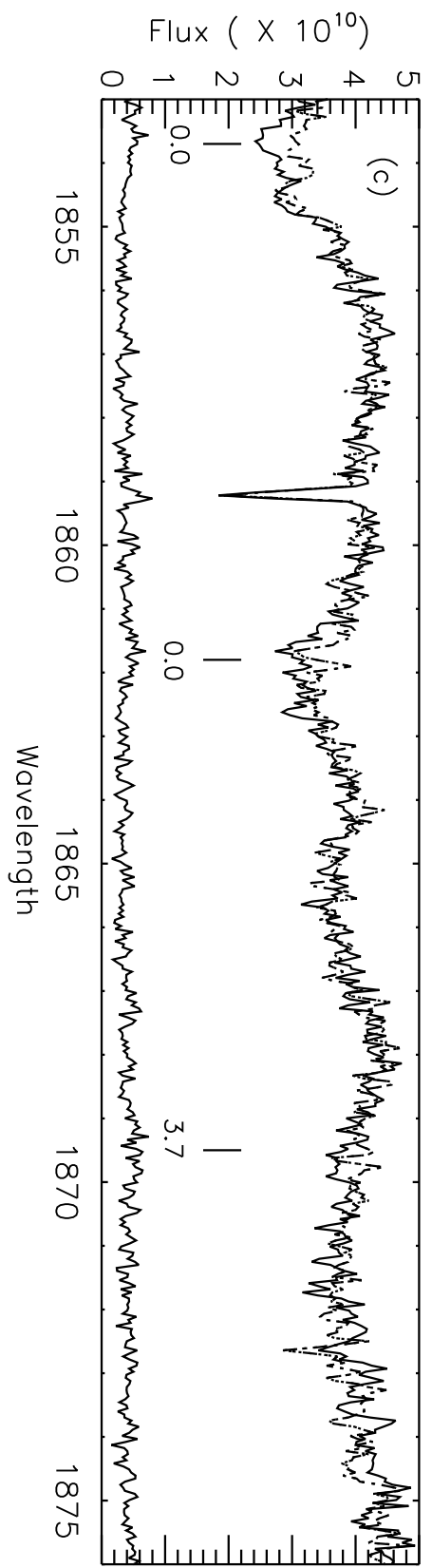
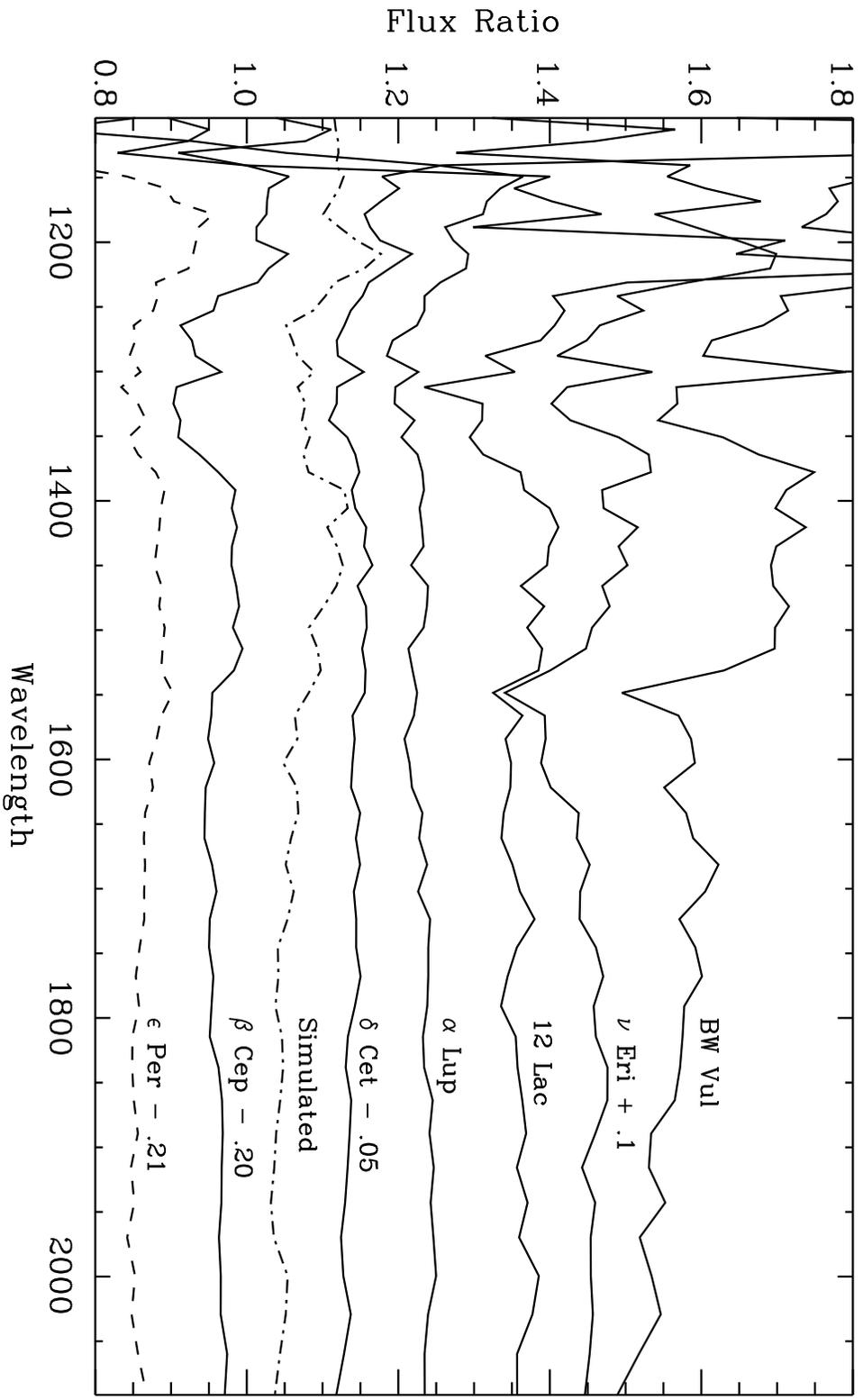
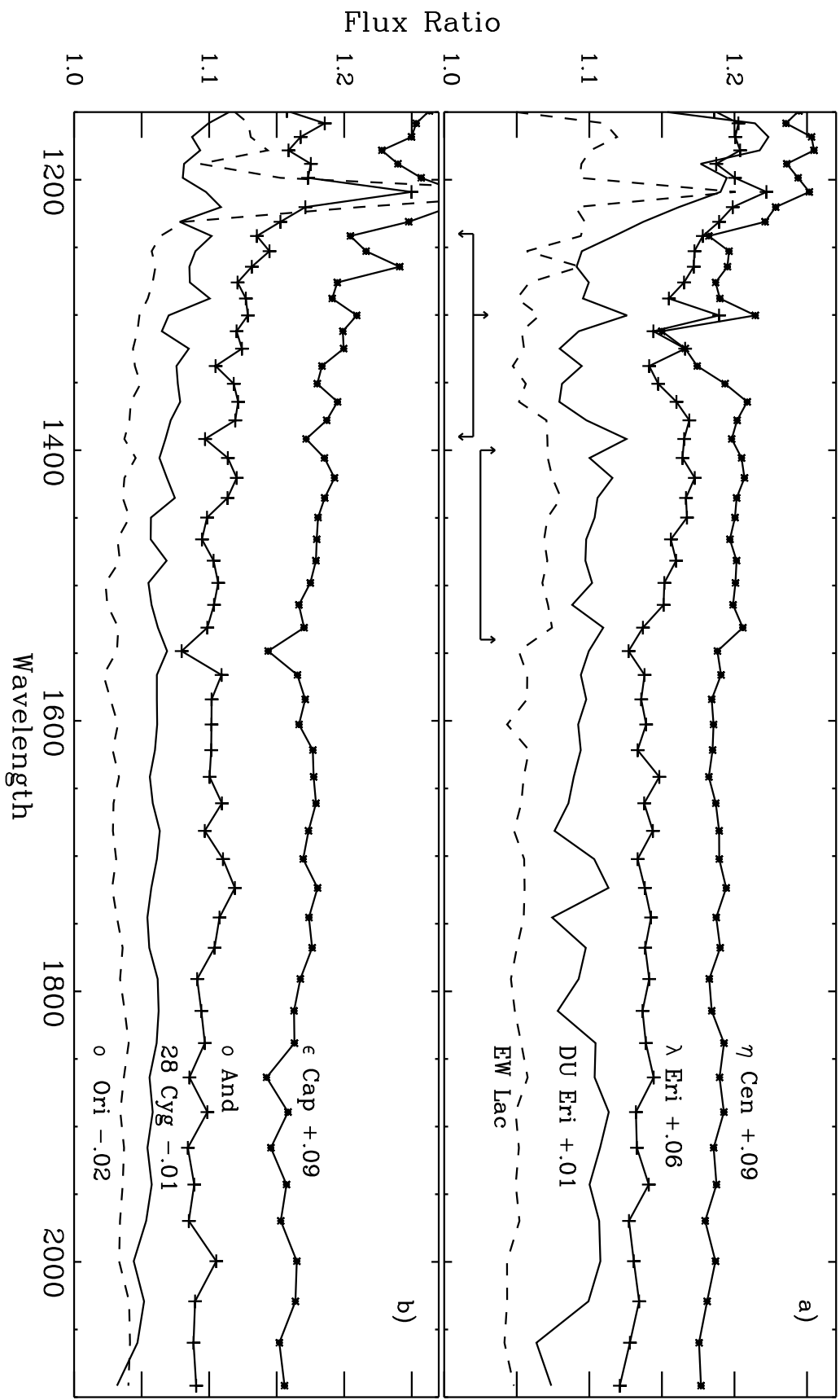
\end{document}